\documentclass[final,3p,12pt]{elsarticle}
\usepackage{amsfonts}
\usepackage{amssymb}
\usepackage{amsmath}
\usepackage{graphicx}
\usepackage[usenames, dvipsnames]{color}
\usepackage{times}

\journal{Physics Letters A}

\begin{document}

\begin{frontmatter}

\title{Stabilization of the Peregrine soliton and Kuznetsov-Ma breathers by means of nonlinearity and dispersion management}

\author{J. Cuevas-Maraver}
\address{Grupo de F\'{i}sica No Lineal, Departamento de F\'{i}sica Aplicada I,
Universidad de Sevilla. Escuela Polit\'{e}cnica Superior, C/ Virgen de \'{A}frica, 7, 41011-Sevilla, Spain \\
Instituto de Matem\'{a}ticas de la Universidad de Sevilla (IMUS). Edificio Celestino Mutis. Avda. Reina Mercedes s/n, 41012-Sevilla, Spain}

\author{Boris A. Malomed}
\address{
Department of Physical Electronics, School of Electrical Engineering,
Faculty of Engineering, \\
Tel Aviv University, Tel Aviv 69978, Israel}

\author{P. G. Kevrekidis}
\address{Department of of Mathematics and Statistics, University of
Massachusetts,\\
Amherst, MA 01003-9305, USA}

\author{D. J. Frantzeskakis}
\address{Department of Physics, National and Kapodistrian University of Athens, Panepistimiopolis,
Zografos, Athens 15784, Greece}

\begin{abstract}
  We demonstrate a possibility to
  make rogue waves (RWs)
in the form of the Peregrine soliton (PS) and
Kuznetsov-Ma breathers (KMBs)
effectively stable objects, with the help of properly defined
dispersion or nonlinearity management applied to the continuous-wave (CW) background
supporting the RWs. 
In particular, it is found that either management scheme,
if applied along the longitudinal coordinate, making the underlying
nonlinear Schr\"{o}dinger equation (NLSE) self-defocusing in the course of
disappearance of the PS, indeed stabilizes the global solution with
respect to the modulational instability of the background.
In the process, additional excitations are generated,
namely, dispersive shock waves and, in some cases, also a pair of
slowly separating dark solitons. Further, the nonlinearity-management
format, which makes the NLSE defocusing outside of a finite domain in
the transverse direction, enables the stabilization of the KMBs,
in the form of confined oscillating states. On the other hand, a
nonlinearity-management format applied periodically along the
propagation direction, creates expanding patterns featuring
multiplication of KMBs through their cascading fission.
\end{abstract}


\end{frontmatter}

\section{Introduction}

The nonlinear Schr\"{o}dinger equation (NLSE) and its variants are well known
as universal models for nonlinear waves and solitons, as well as relevant phenomenology,
in many areas of physics including water
waves, plasmas, nonlinear optics, Bose-Einstein condensates, and so on.
\cite{NLSE1,Agrawal,NLSE2,NLSE3,NLSE4,Barcelona,NLSE5,Boris,Mihalache}. Among various 
solutions of these equations, a class of unstable but physically meaningful
ones represent rogue waves
(RWs), which can spontaneously emerge on top of continuous-wave (CW)
modulationally (alias Benjamin-Feir \cite{BF,MI}) unstable states, and
then disappear. RWs were originally identified in terms of water waves in
the ocean \cite{Pelinovsky}. Later, this concept was extended to nonlinear
fiber optics \cite{Nature,NaturePhot,Turitsyn,Dudley,suret}
and other areas (see, e.g., Refs.~\cite{Bludov,Taki,RW1,Zhenya}).
Recently, the pioneering work of~\cite{bertola} argued that the
so-called Peregrine solitons (PSs) are a generic byproduct of
a phenomenon called gradient catastrophe arising at the level of the
semi-classical form of the NLSE. Moreover such solutions also emerged
in the context of interactions of dispersive shock waves~\cite{khamis}.
An overview of the current state of the
studies of RWs can be found in Ref.~\cite{RW2,Chen}.

The classical integrable NLSE with the cubic self-focusing nonlinearity, in
terms of the spatial-domain propagation (or with the anomalous
group-velocity dispersion (GVD), in terms of fiber optics \cite{Agrawal})
gives rise both to the CW states subject to the modulational instability,
and to exact RW solutions, the most fundamental ones being the Peregrine
soliton (PS) \cite{Peregrine}, the Kuznetsov-Ma breather (KMB) \cite{Kuzn,Ma},
and the Akhmediev breather \cite{Akhm}.
The PS is a state of an instanton type built on top of the CW background,
i.e., it is localized both in the longitudinal
and transverse coordinates (if the NLSE is considered as a model of a planar
waveguide in the spatial domain).
The KMB, on the other hand, is localized in the transverse direction,
and periodically oscillate in the longitudinal one, while the Akhmediev breather \cite{Akhm},
is periodic in the transverse direction and self-localized along the
propagation distance. Due to the fact that all these states are supported by
the modulationally unstable background, they are unstable too, which poses a
limitation to their physical realizations; even when they
are carefully realized experimentally~\cite{suret}, the
modulational instability of the background cannot be avoided.
On the other hand, the concept of the
dispersion and nonlinearity management \cite{book,Barcelona} suggests a
possibility to stabilize RWs by making the GVD and/or local nonlinearity
coefficients functions of the propagation distance or transverse coordinate.
This way, these solitons and breathers would have enough room to emerge in
areas where the NLSE is self-focusing, and, on the other hand, the
background may be globally stabilized by making the NLSE self-defocusing
outside of the area reserved for the formation of the RWs. The objective of
the present work is to demonstrate the ``proof of principle'' as regards
these possibilities for the effective
stabilization of the PS and KMBs, applying the schemes of both the
dispersion and nonlinearity management. While our focus here is on
numerical experiments, the existence~\cite{book} and earlier experimental
implementation~\cite{book,psaltis} of related schemes suggests their
potential consideration  in (near-)future optical and related physical systems.

The paper is organized as follows. The model and numerical
methods used for its analysis are presented in Section II. The results
obtained for the stabilization of the PS and KMBs, under the action of the management,
are reported, respectively, in Sections III and IV (while both
the dispersion and nonlinearity management are applied to the PS, only the
latter scheme is considered for the KMBs). Finally, the paper is concluded by Section~V.

\section{The model and numerical scheme}

The NLSE which we use for the stabilization of the PSs and KMBs is taken as

\begin{equation}
iu_{z}+\frac{1}{2}D(z)u_{xx}+\gamma (x,z)|u|^{2}u=0.  \label{eq:NLS}
\end{equation}%
In the spatial domain, which corresponds to the light propagation in a
planar waveguide, the diffraction coefficient is constant, $D(z)\equiv 1$,
while the local nonlinearity coefficient may be modulated as a function of
the propagation and transverse coordinates, $z$ and $x$ \cite{Barcelona}. In
the temporal domain, corresponding to the light propagation in an optical
fiber, $x$ is actually the reduced time, $\tau \equiv t-z/V_{\mathrm{gr}}$ ($%
t$ is time proper, and $V_{\mathrm{gr}}$ is the group velocity of the
carrier wave), the relevant fiber's model has $\gamma (x,z)\equiv 1$, while
the GVD\ coefficient, $D(z)$ may be made a function of the propagation
length, using known techniques of the GVD management \cite{book,Turitsyn}.

The integrable version of the NLSE, i.e., Eq. (\ref{eq:NLS}) with $%
D(z)\equiv 1$ and $\gamma (x,z)\equiv 1$, gives rise to the exact PS \cite%
{Peregrine} and KMB \cite{Kuzn,Ma} solutions:

\begin{equation}
u_{\mathrm{PS}}(x,z)=\left[ 1-\frac{4(1+2\mathrm{i}z)}{1+4x^{2}+4z^{2}}%
\right] \mathrm{e}^{\mathrm{i}z}.  \label{eq:PS}
\end{equation}

\begin{equation}
u_{\mathrm{KMB}}(x,z)=\left[ 1+\frac{2(1-2a)\cos (\omega z)-\mathrm{i}\omega
\sin (\omega z)}{\sqrt{2a}\cosh (bx)-\cos (\omega z)}\right] \mathrm{e}^{%
\mathrm{i}z},  \label{eq:KMB}
\end{equation}%
where $a\equiv \left( 1+\sqrt{\omega ^{2}+1}\right) /4$ and $b\equiv 2\sqrt{%
2a-1}$, while $\omega $ is an arbitrary frequency of the KMB oscillations.
As explained in the Introduction, both solutions are supported by the CW
background, $\exp(iz)$, which is prone to the modulational instability.

To demonstrate effects of management, we present here results of numerical
simulations of Eq. (\ref{eq:NLS})\ with initial condition:
\begin{equation}
u(x,0)=u_{\mathrm{PS}}(x,z_{0}),~z_{0}=-5,  \label{PS}
\end{equation}
when dealing with PS  (the choice of $z_{0}=-5$ is appropriate for
demonstrating both the growth and the decay phase of the
wave structure). In the case of KMBs, the input is
taken as:
\begin{equation}
u(x,0)=u_{\mathrm{KMB}}(x,0).  \label{KMB}
\end{equation}
In the latter case, we set $\omega =1.5$ here, as this value was found to be
appropriate for representing the generic situation. Note that, as RW
solutions possess relatively steep peaks, the present version of the NLSE is
a mildly stiff equation for simulations, in these cases. To handle it, we
have used the exponential time differencing fourth-order Runge-Kutta
numerical algorithm~\cite{Kassam}. The discretization of the second derivative was
performed by dint of the Fourier spectral collocation,
implying periodic boundary conditions imposed on the integration domain, $-L<x<+L$.
Here we report results produced for $L=200$, and a discretization
spacing $\Delta x=25/256\approx 0.10$, as well as a time step $\Delta t=(\Delta
x)^{2}/4$. These parameters ensure the stability of the numerical
integration.

\begin{figure}[tbp]
\centering
\begin{tabular}{cc}
\includegraphics[width=4.5cm]{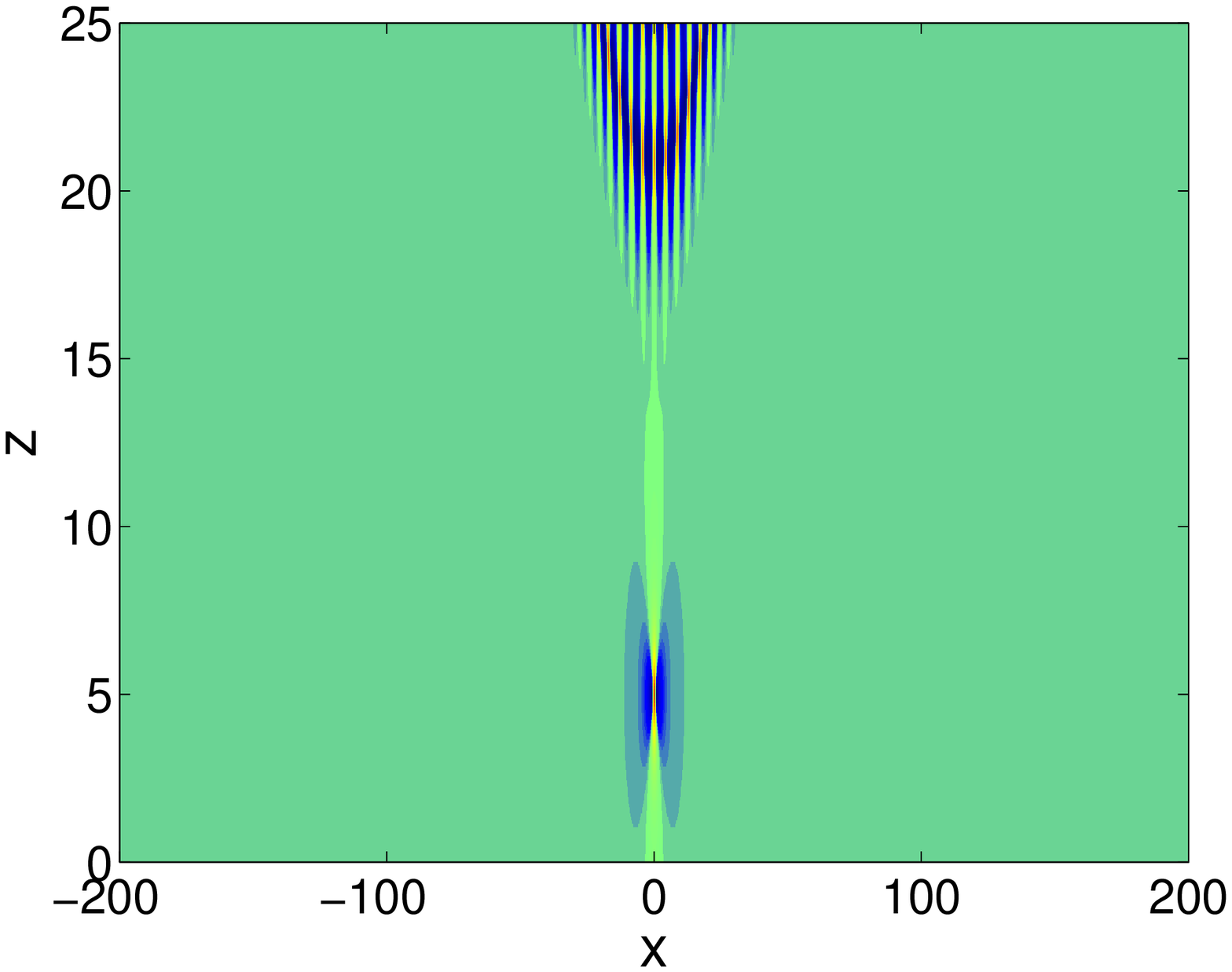} & %
\includegraphics[width=4.5cm]{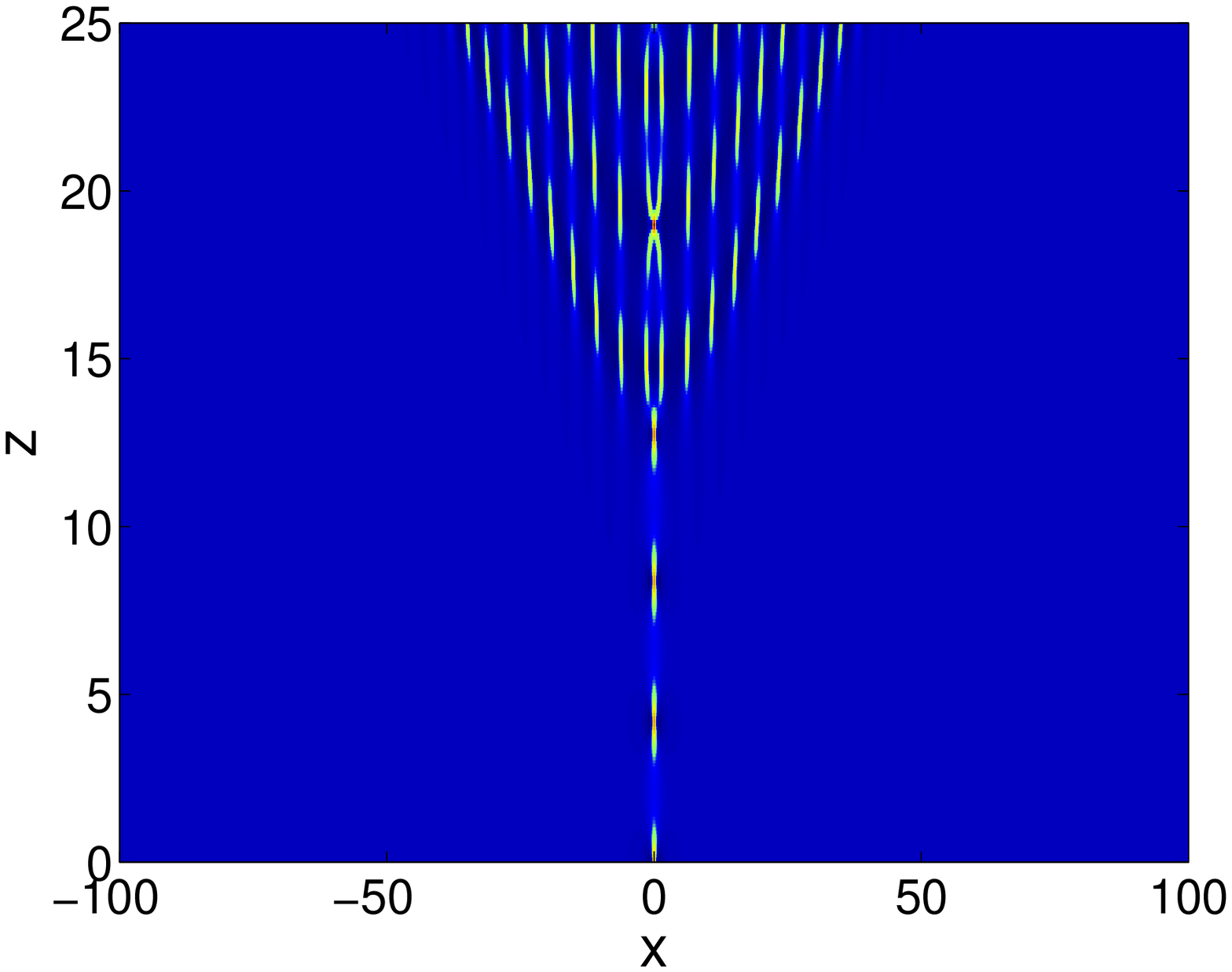}%
\end{tabular}%
\caption{Density plots illustrating the evolution of a Peregrine soliton
(left) and a Kuznetsov-Ma breather with $\protect\omega =1.5$ (right) in the
framework of the constant-coefficient NLSE (\protect\ref{eq:NLS}), which
does not include any management.}
\label{fig:noman}
\end{figure}

Figure~\ref{fig:noman} shows the outcome of the simulations performed for
the NLSE (\ref{eq:NLS}) in the absence of management, $D=\gamma \equiv 1$,
using the above-mentioned PS and KMB wave forms as initial conditions. The
onset of the modulational instability, seeded by truncation errors of the
numerical algorithm, is clearly observed at the center of the domain.
It is natural that this occurs there, as the presence of the PS
amplifies growing perturbations on top of the unstable background.
Notice that, recently, the instability of the KMB --and by extension of the PS in the limit
of vanishing frequency-- was analyzed via Floquet theory in Ref.~\cite{jcuevas}.

\section{The management of Peregrine solitons}

First, we test the effects of the management applied to the PS. For this
purpose, we have performed simulations of Eq.~(\ref{eq:NLS}) with either $%
D\equiv 1$ and $z$-dependent nonlinearity $\gamma (z)$, or vice versa. As we
show below, in both cases outcomes are quite similar. 
The nonlinearity management is implemented as:
\begin{equation}
\begin{split}
\gamma (x,z)& =\left\{
\begin{array}{ll}
1 & \text{at }\ z<z_{1} \\
-1 & \text{at }\ z\geq z_{1}%
\end{array}%
,\right.  \\
D& \equiv 1,
\end{split}
\label{eq:gammaPS}
\end{equation}
i.e., the originally focusing nonlinearity switches to defocusing at $z=z_{1}$,
while the dispersion management can be introduced as
\begin{equation}
\begin{split}
D(z)& =\left\{
\begin{array}{ll}
1 & \text{for }\ z<z_{1} \\
-1 & \text{for }\ z\geq z_{1}%
\end{array}%
,\right.  \\
\gamma (x,z)& \equiv 1.
\end{split}
\label{eq:dispPS}
\end{equation}
In the latter case, the nonlinearity keeps the focusing sign, while the GVD
changes from anomalous to normal at $z=z_{1}$. As said above, the results
shown here correspond to the PS launched by means of input~(\ref{PS}).

\begin{figure}[tbp]
\centering
\begin{tabular}{ccc}
\includegraphics[width=4.5cm]{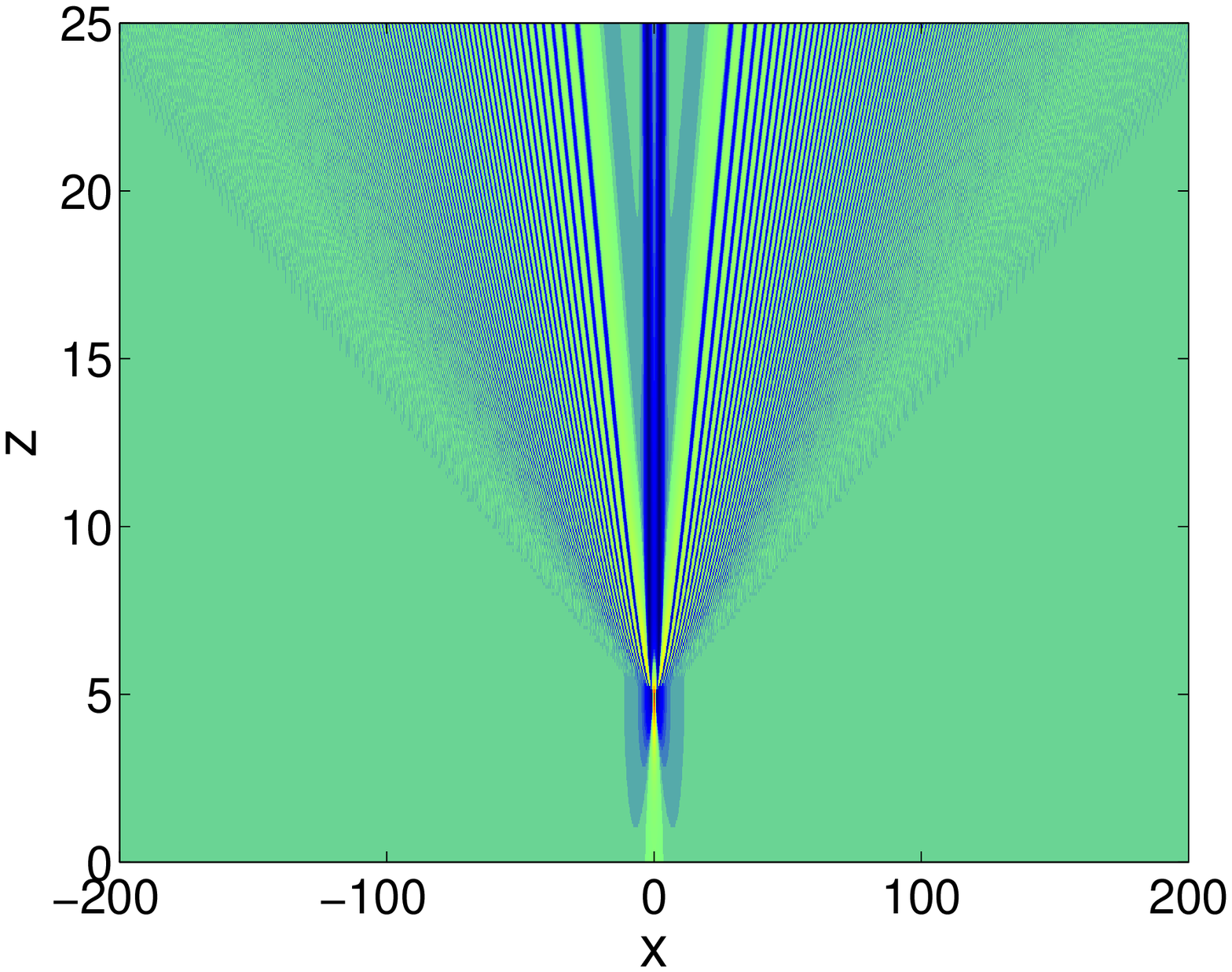} & %
\includegraphics[width=4.5cm]{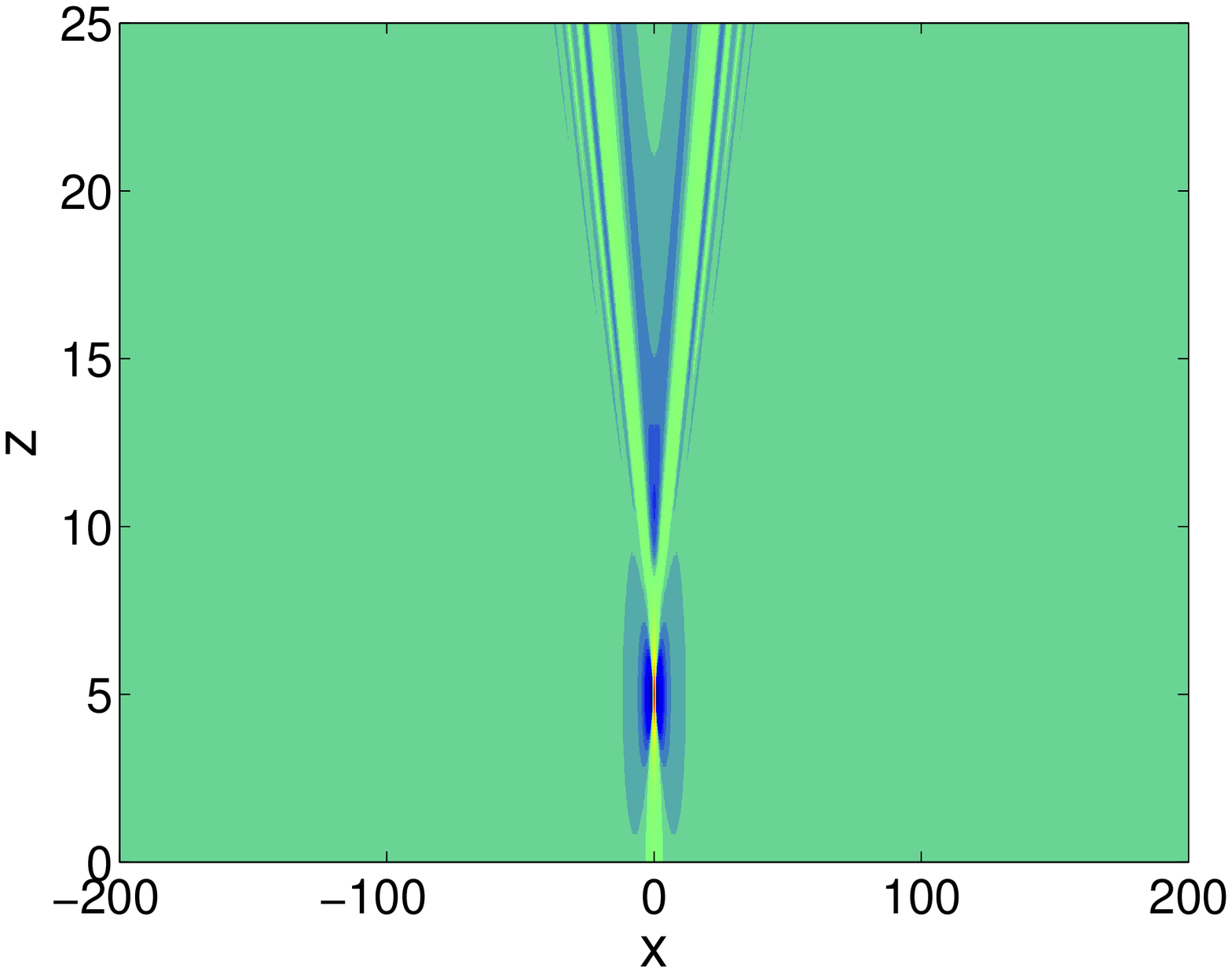} & %
\includegraphics[width=4.5cm]{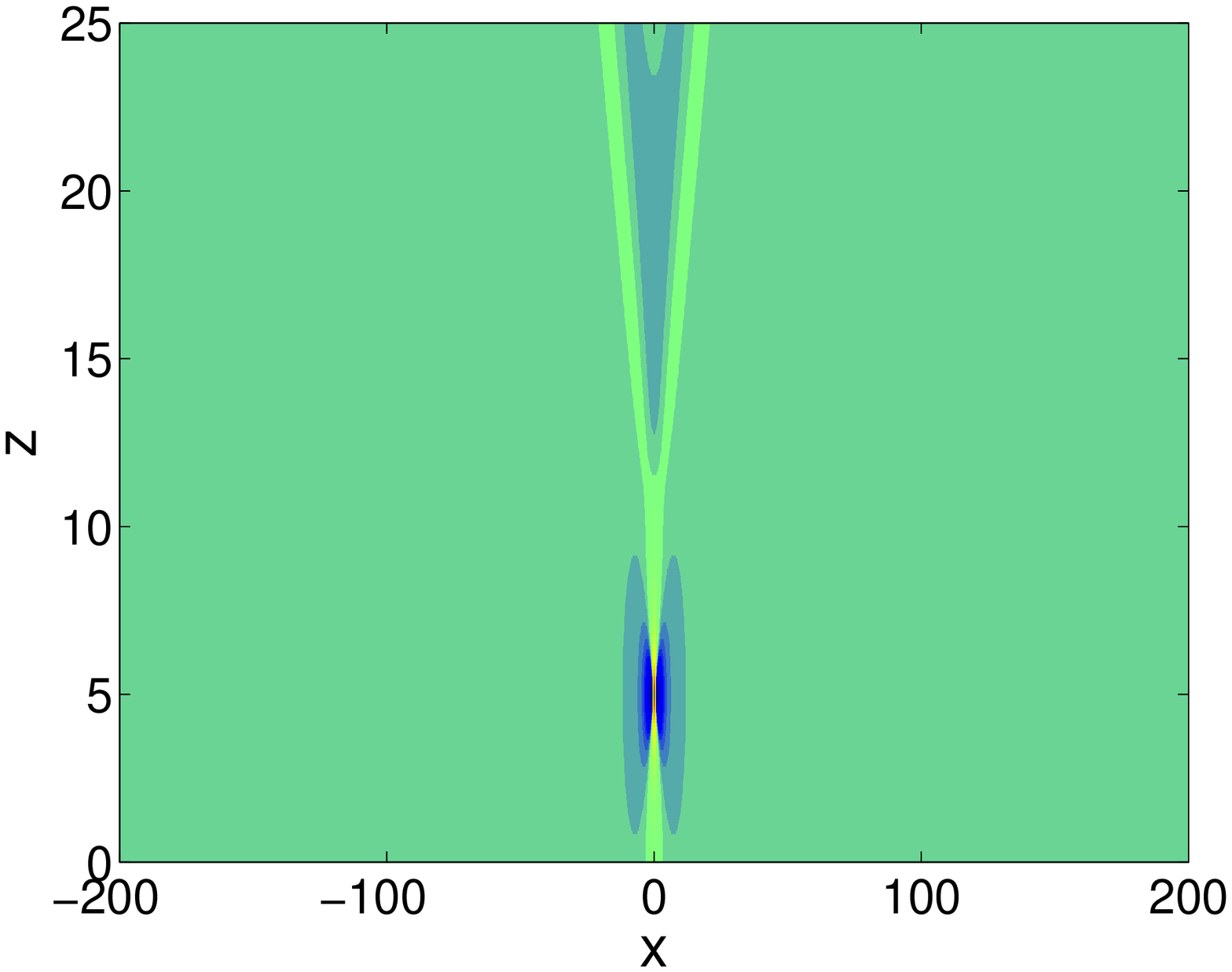} \\
\includegraphics[width=4.5cm]{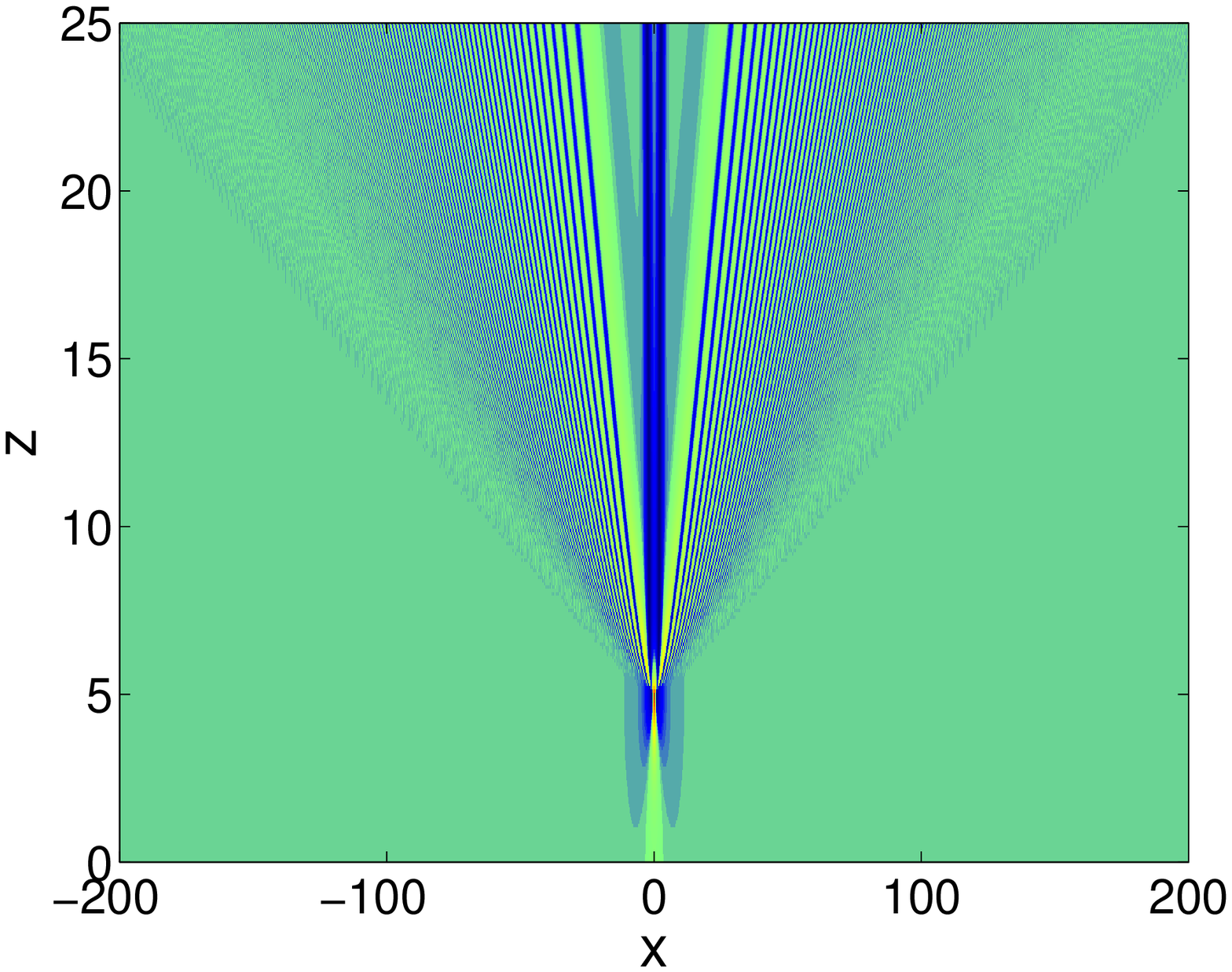} & %
\includegraphics[width=4.5cm]{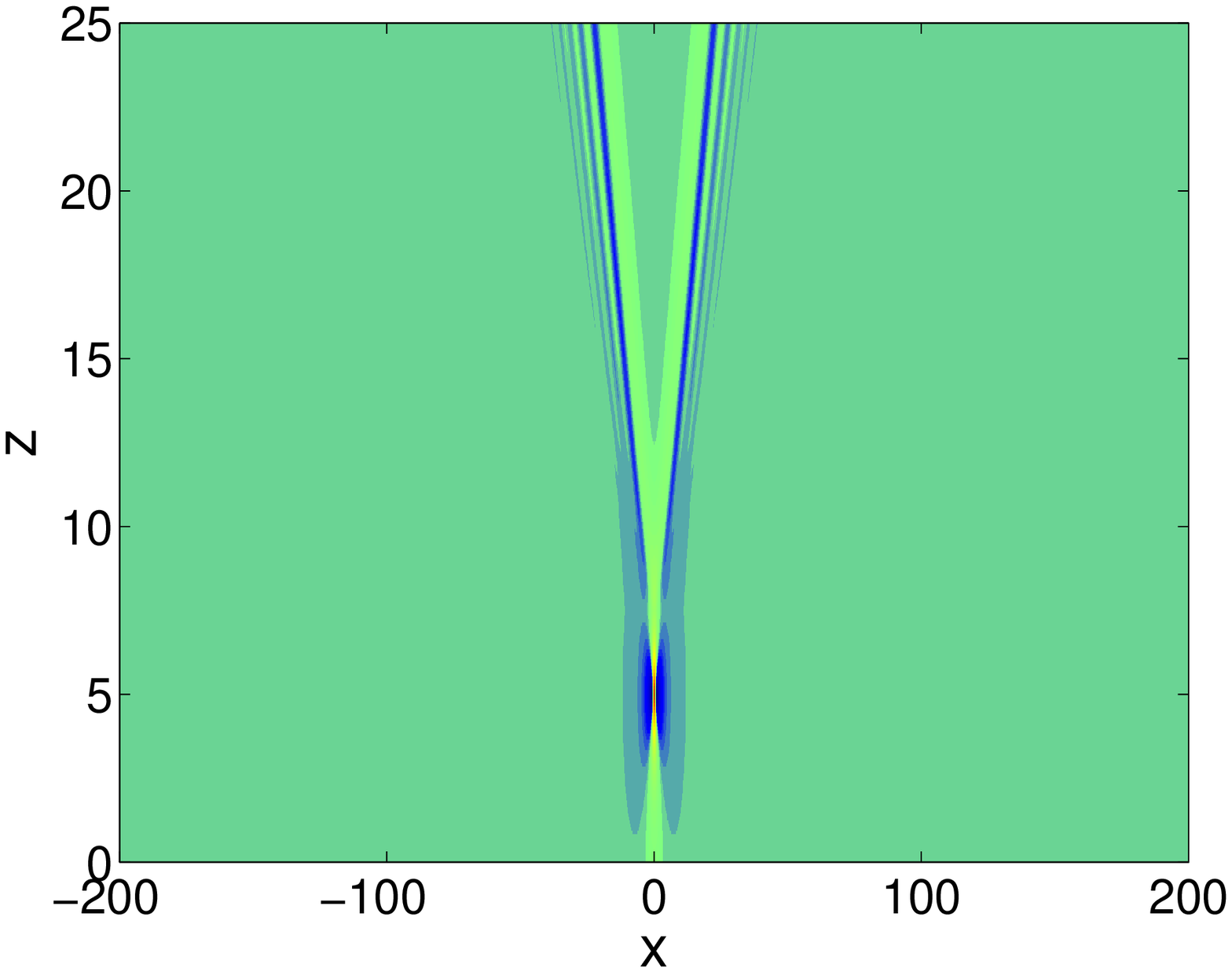} & %
\includegraphics[width=4.5cm]{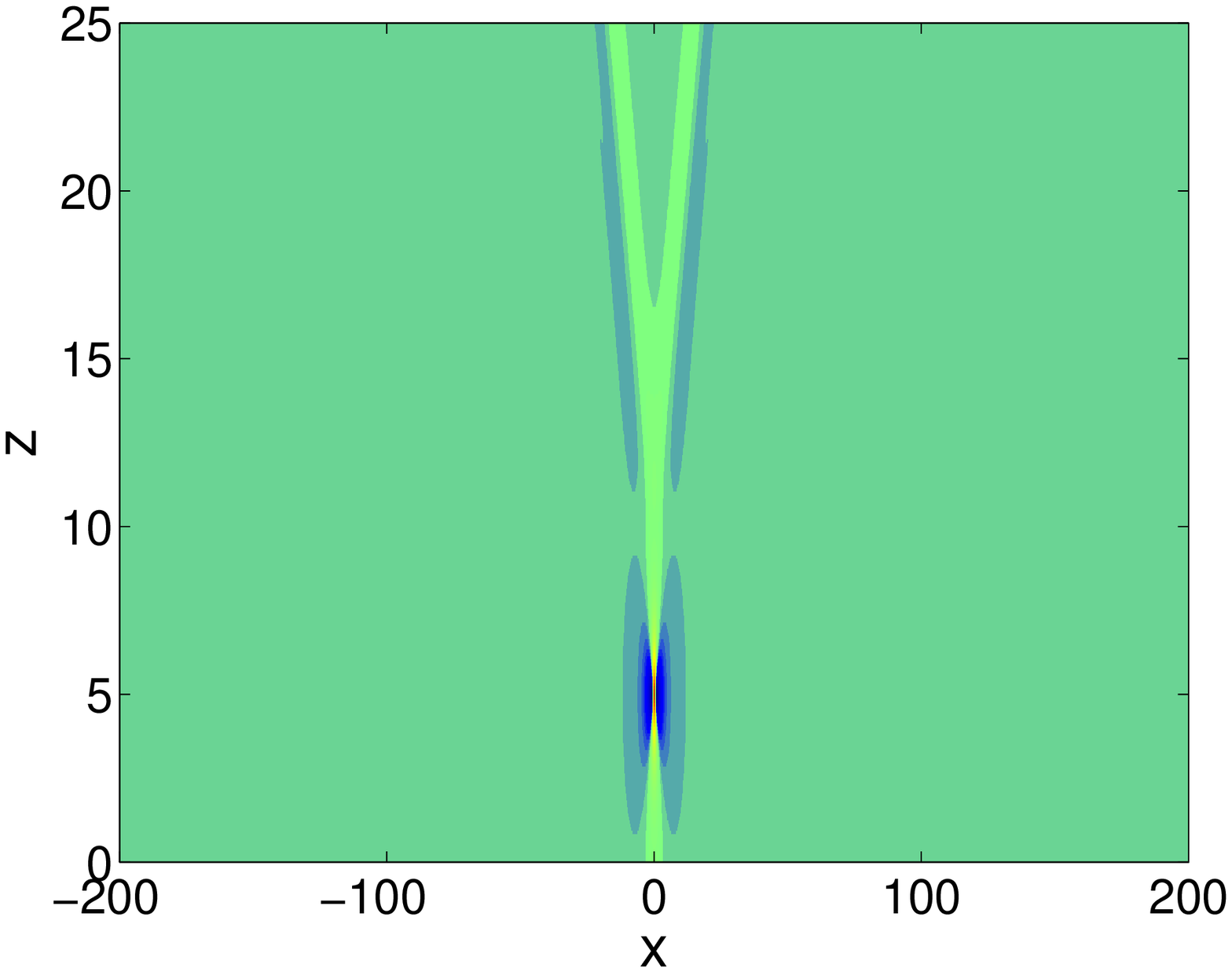} \\
\includegraphics[width=4.5cm]{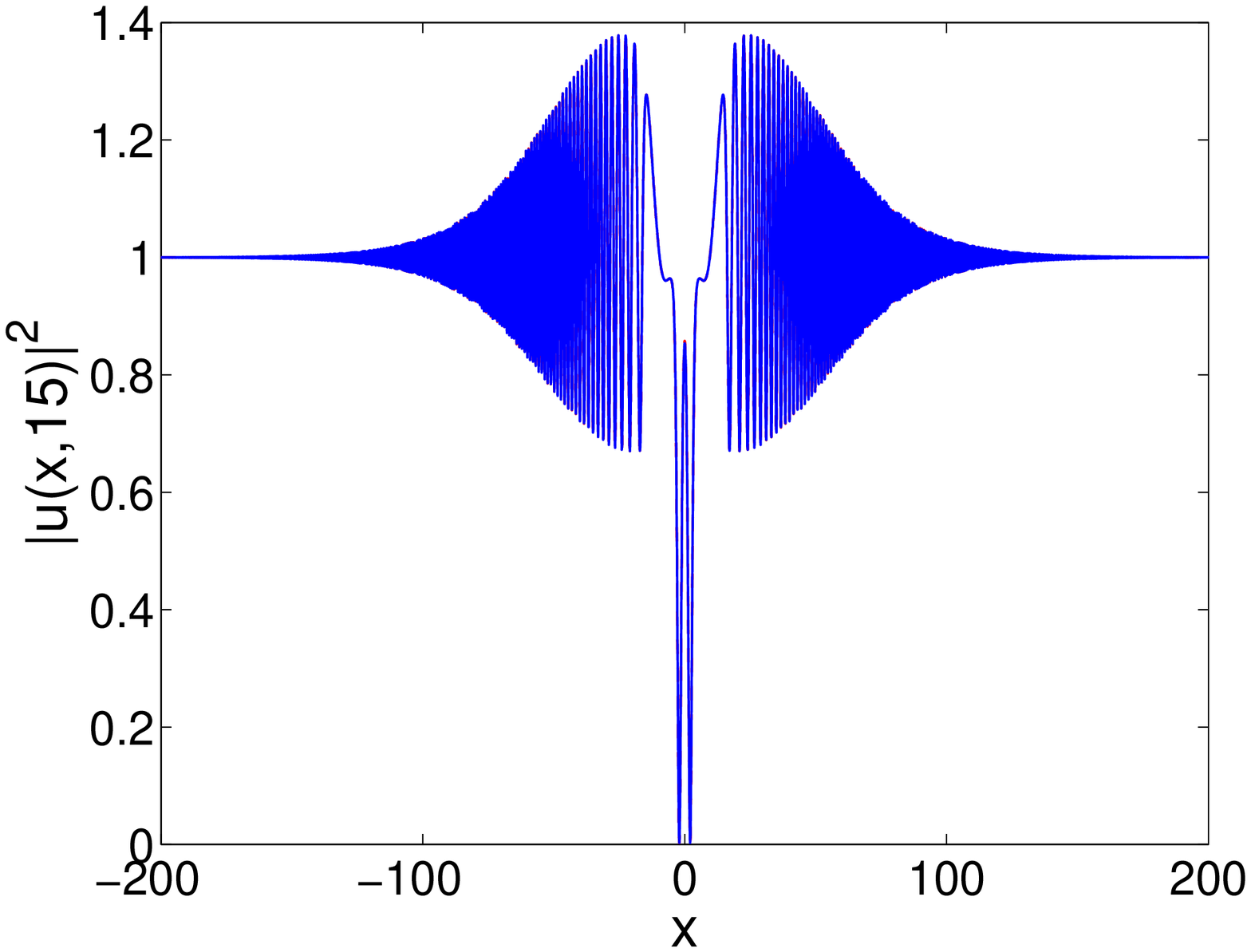} & %
\includegraphics[width=4.5cm]{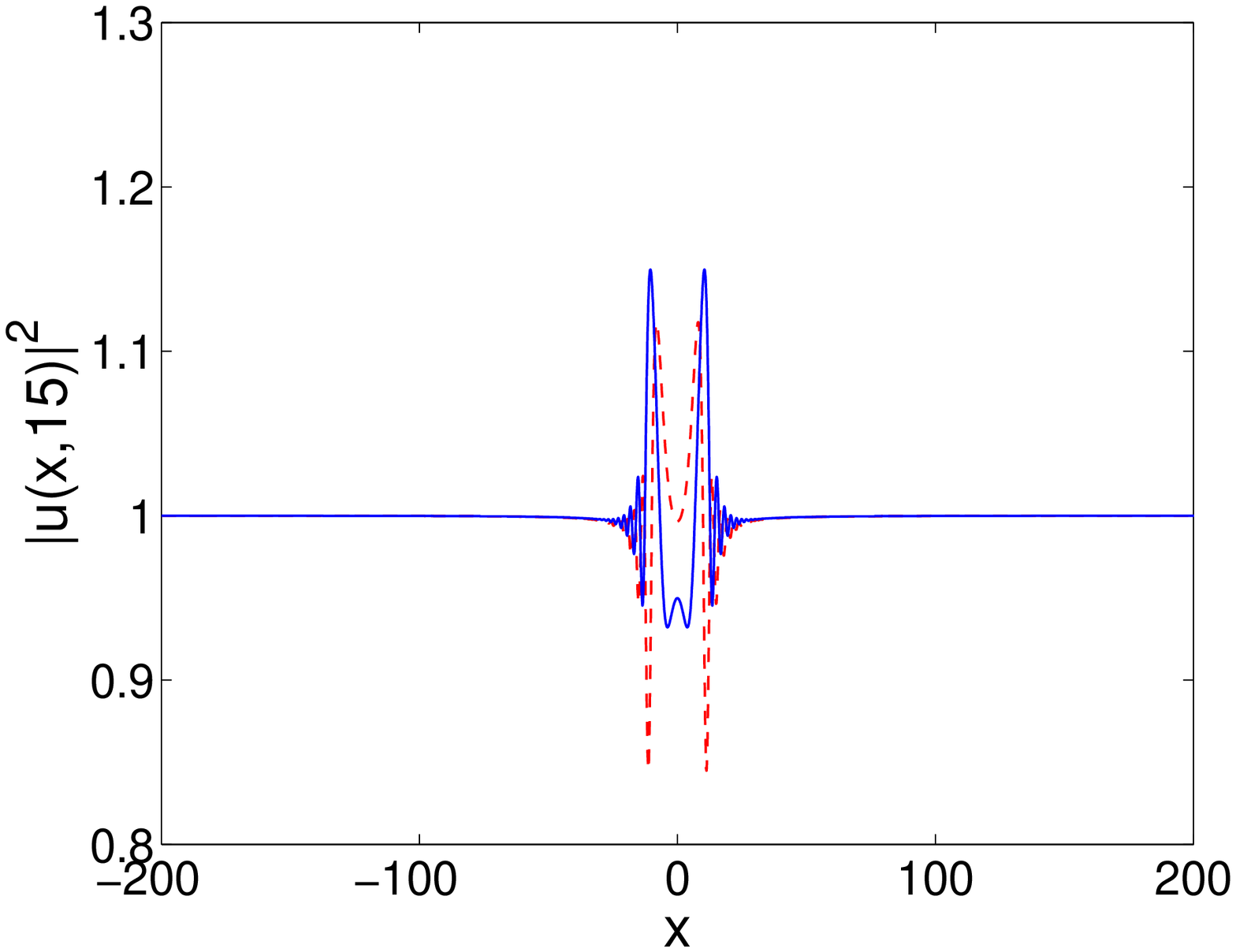} & %
\includegraphics[width=4.5cm]{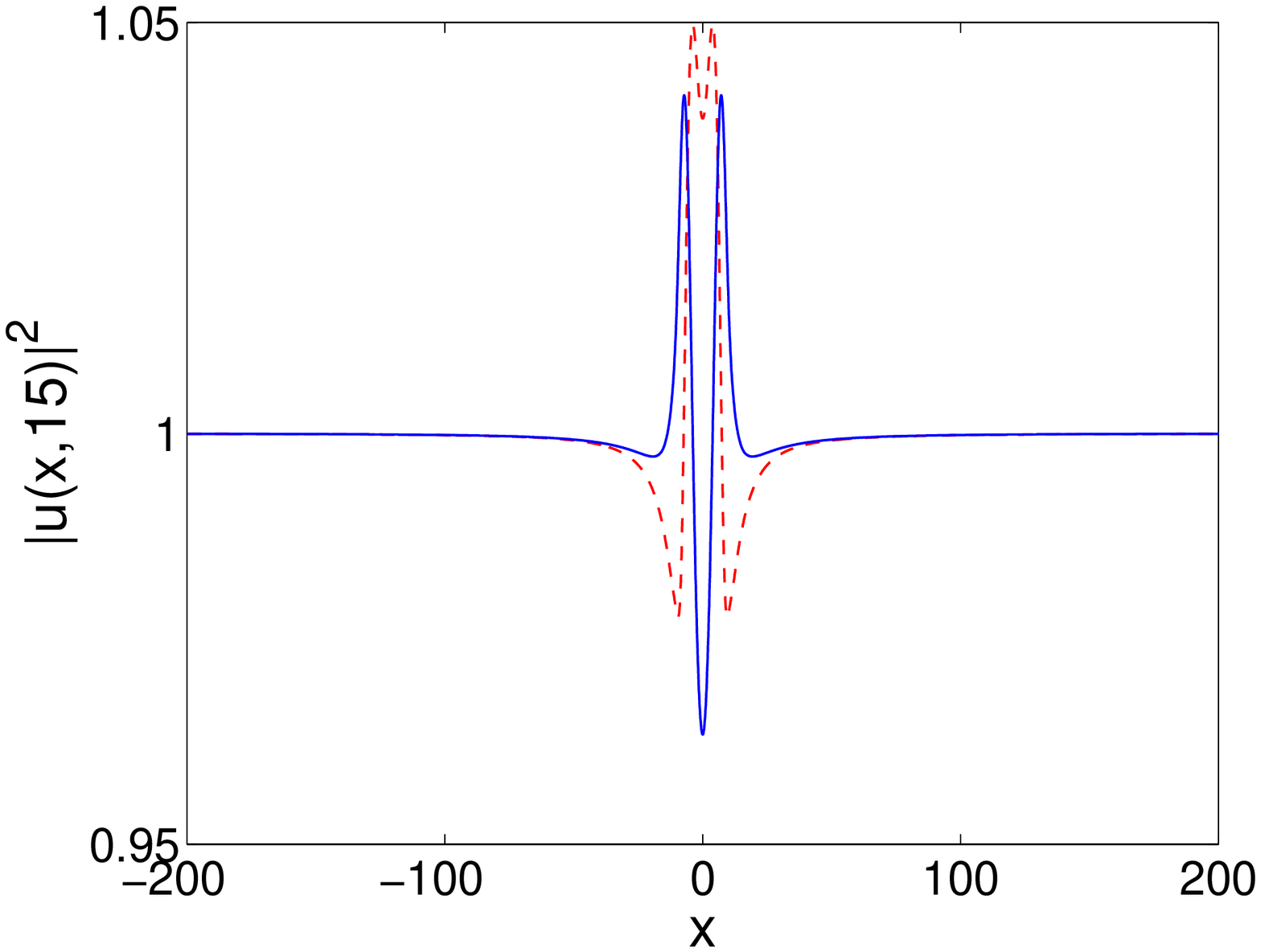} \\
\includegraphics[width=4.5cm]{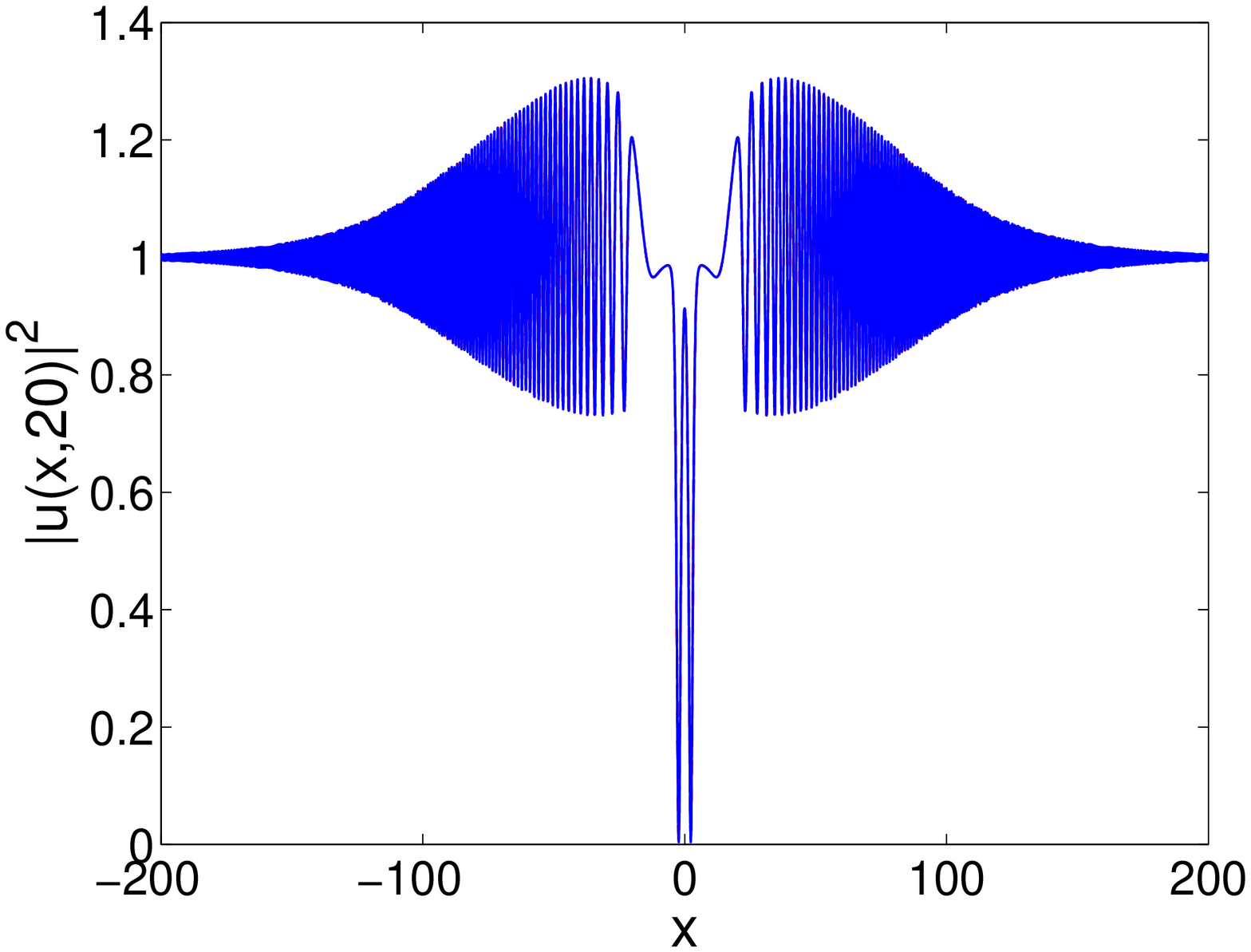} & %
\includegraphics[width=4.5cm]{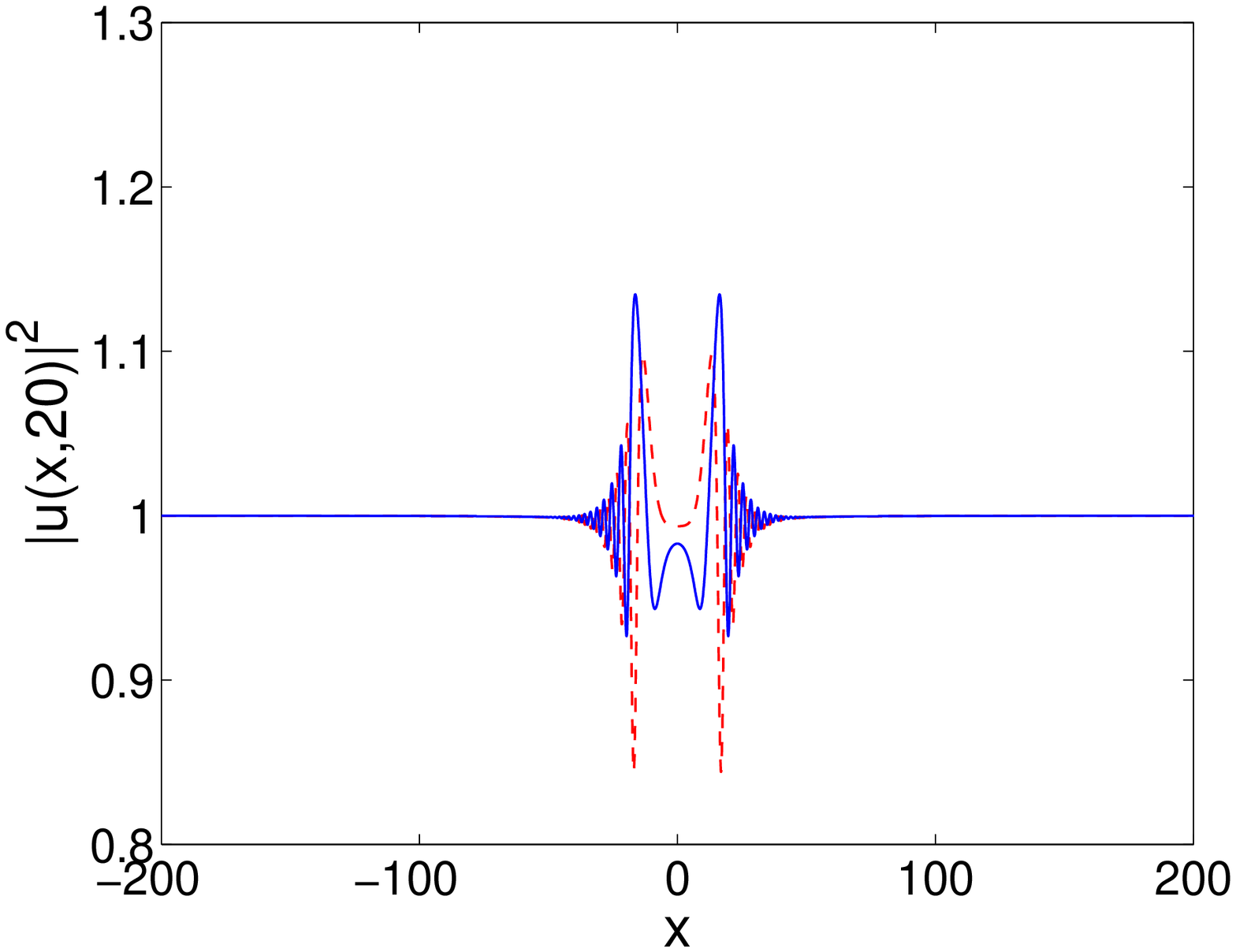} & %
\includegraphics[width=4.5cm]{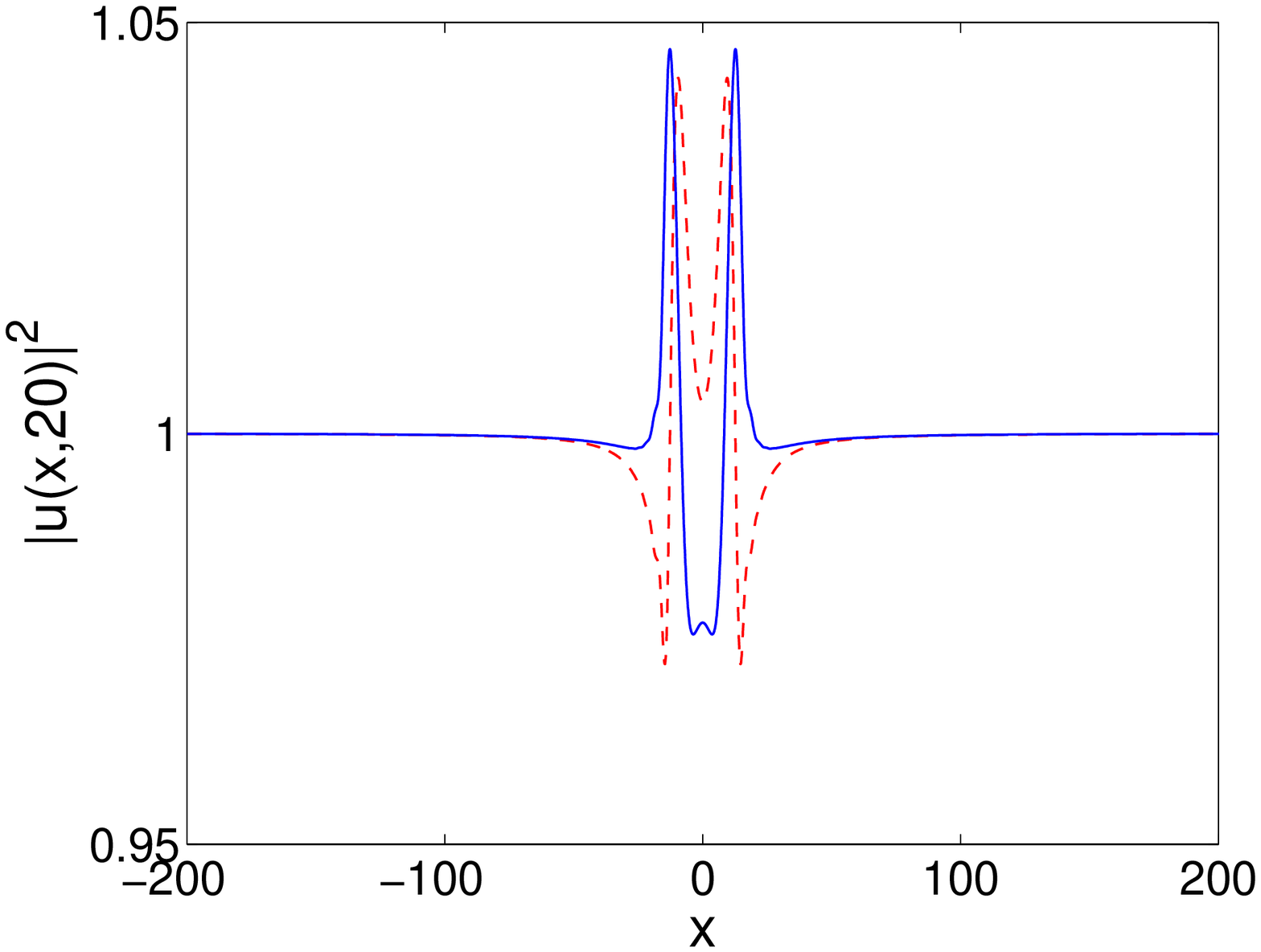} \\
\includegraphics[width=4.5cm]{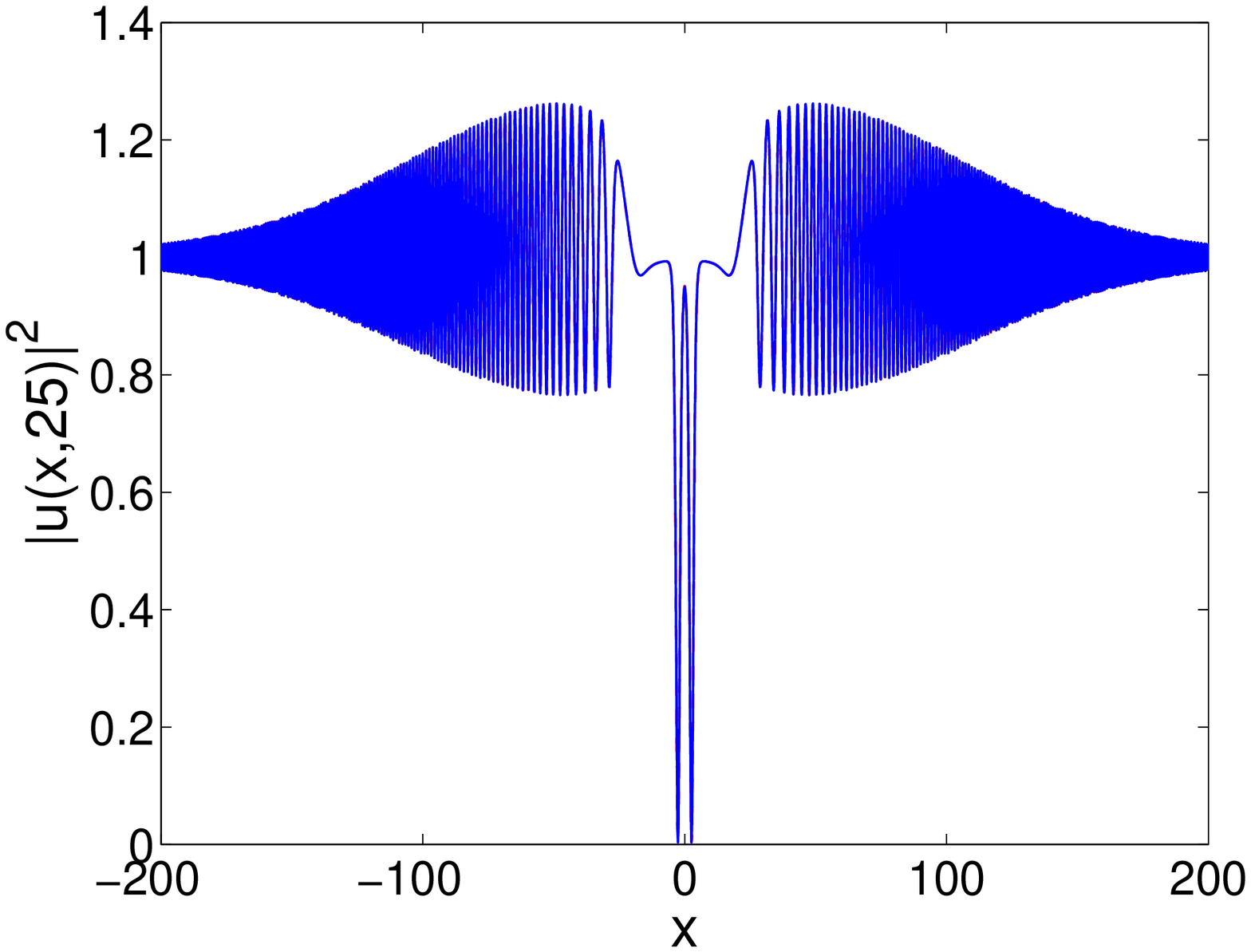} & %
\includegraphics[width=4.5cm]{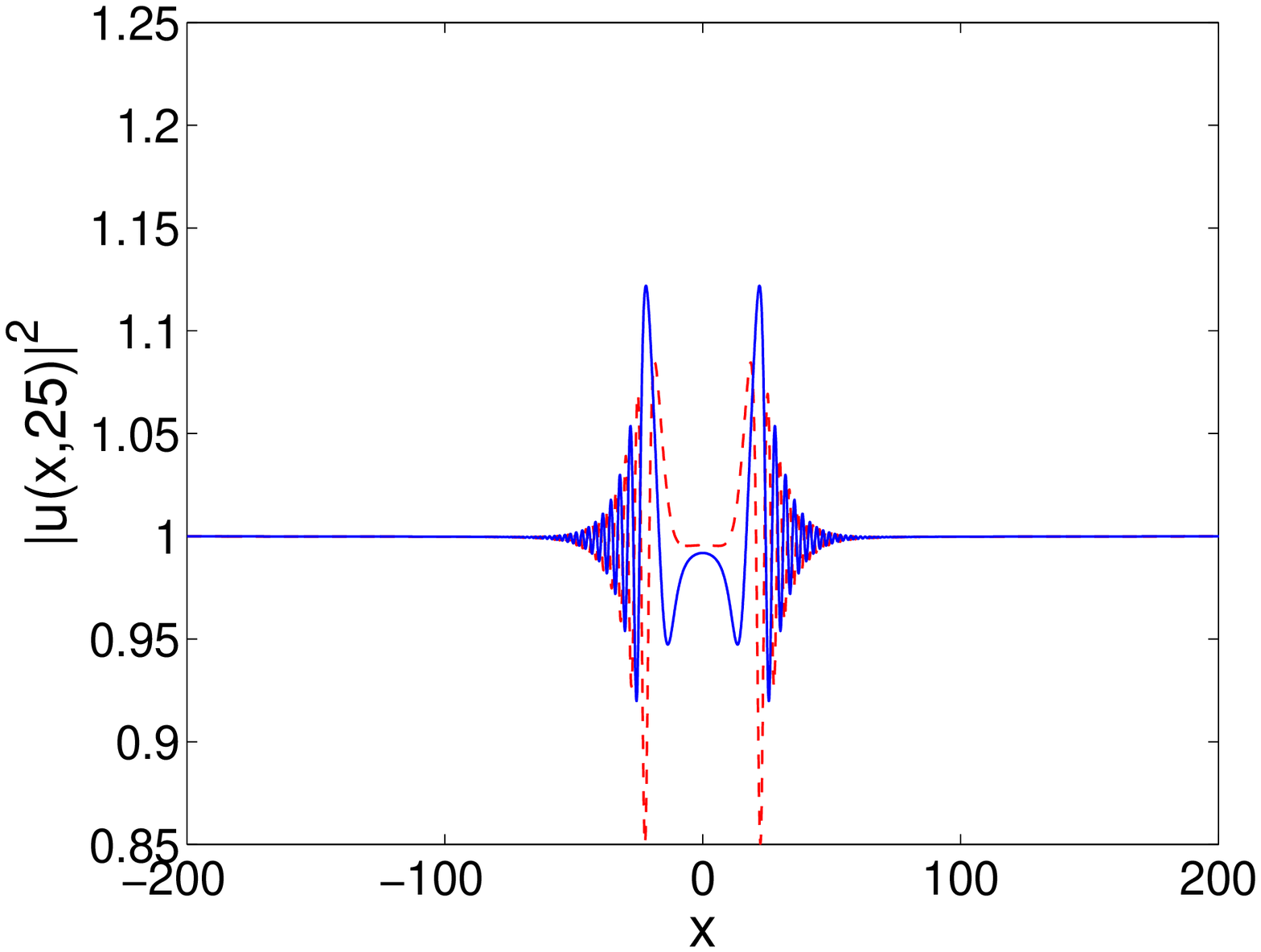} & %
\includegraphics[width=4.5cm]{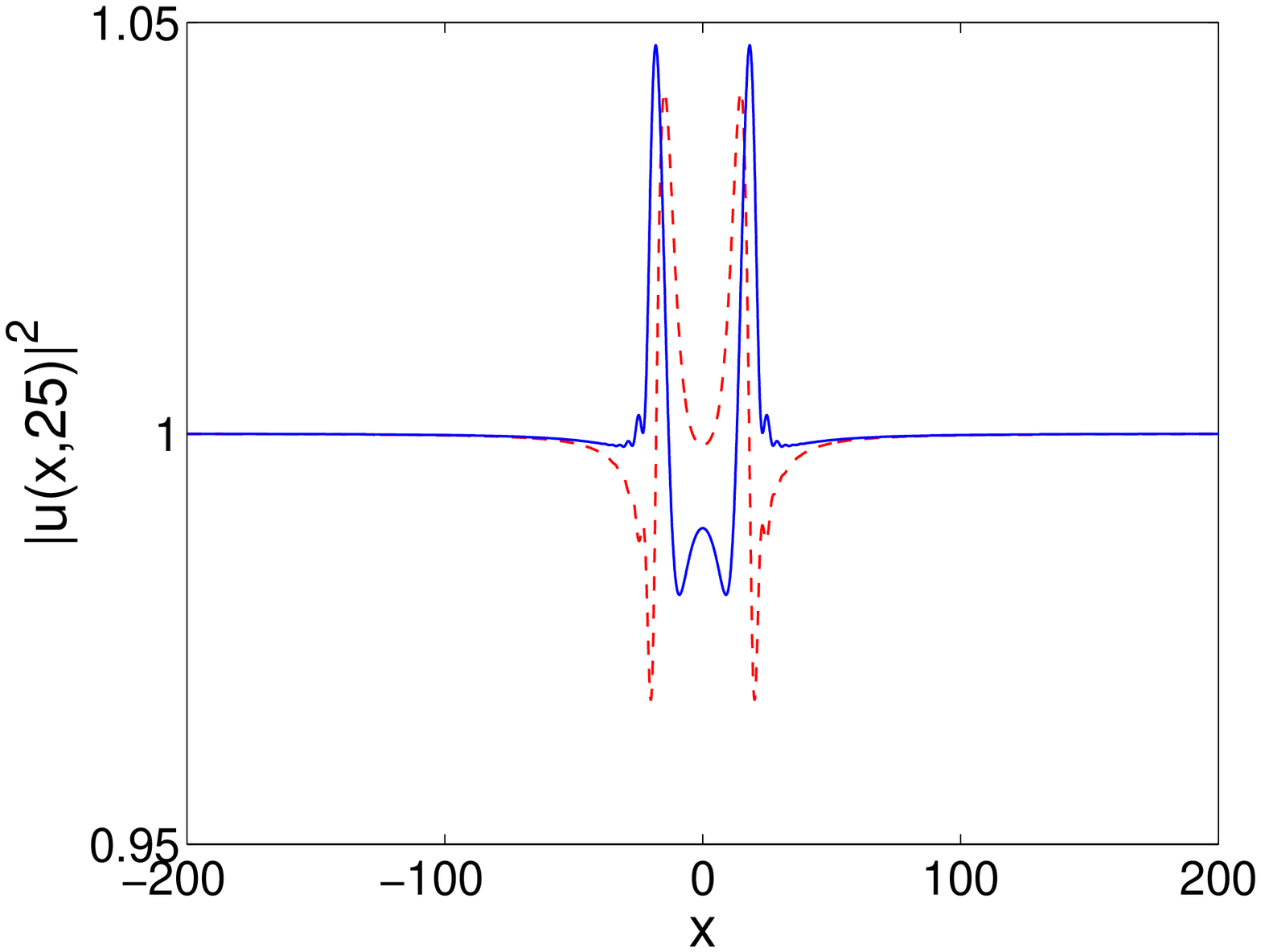} \\
&  &
\end{tabular}%
\caption{Evolution of Peregrine solitons under the action of the
management, with $z_{1}=-z_{0}=5$ (left panels), $z_{1}=7.5$ (central
panels) and $z_{1}=10$ (right panels). Panels in the first and second row
display density plots under the action of the nonlinearity and
dispersion management, see Eqs. (\protect\ref{eq:gammaPS}) and (\protect\ref%
{eq:dispPS}), respectively. Other panels show snapshots of the managed
Peregrine soliton at different values of $z$, with continuous blue and
dashed red lines corresponding, respectively, to the nonlinearity and
dispersion management.}
\label{fig:PSmanaged}
\end{figure}

Typical examples demonstrating the application of the nonlinearity and
dispersion management to the PS are displayed in Fig.~\ref{fig:PSmanaged}.
It is observed that, as expected, the modulational instability is
suppressed, and the main additional excitations arising past the
disappearance of the PS are the dispersive shock waves
(cf. Refs.~\cite{shock1,shock2,shock3,shock4,shock5,shock6} and for a recent
review Ref.~\cite{shock7}) propagating on top of the uniform background (the
modulationally stable one, due to the adopted management format).
Remarkably, at $z_{1}=-z_{0}=5$, a pair of dark solitons is formed too.
These dark solitons separate slowly because the repulsive interaction between
them is weak, being screened by the shock-wave pattern. For larger values of
$z_{1}$, dark solitons do not emerge; instead, there appear a pair of central dips,
whose depth quickly decreases with $z_{1}$. The depth of the dip is related to
that of the exact PS (\ref{eq:PS}) at $z=z_{1}$.

Finally, we note that similar dynamical scenarios are observed under the
action of the nonlinearity and dispersion managements. Differences between
these two management schemes, which increase with $z_{1}$, amount to
quantitative (yet no major qualitative) details.

\section{The management of Kuznetsov-Ma breathers}

Given the similarity of dispersion and nonlinearity management for the PSs,
in the case of KMBs, we have systematically studied only the nonlinearity
management, fixing $D(z)\equiv 1$ in NLSE\ (\ref{eq:NLS}). We have
considered two different management formats. One of them acts only along the
transverse coordinate, $x$, without dependence on the propagation coordinate
($z$):
\begin{equation}
\gamma (x,z)=\left\{
\begin{array}{ll}
1 & \text{for }\ |x|<x_{1}, \\
-1 & \text{for }\ |x|\geq x_{1},%
\end{array}%
\right.  \label{eq:gammaKM1}
\end{equation}
that is, the nonlinearity is focusing at $|x|<x_{1}$ and defocusing at $|x|\geq
x_{1}$.
The other
format acts periodically along $z$ (in accordance with the fact that the
exact KMB solution (\ref{eq:KMB}) is a periodic function if $z$), being
independent of $x$:

\begin{equation}
\gamma (x,z)=\cos (\omega z).  \label{eq:gammaKM2}
\end{equation}

Generic examples of the numerical results produced by formats (\ref%
{eq:gammaKM1}) and (\ref{eq:gammaKM2}) are displayed in Fig.~\ref%
{fig:KMmanaged}. In the former case, with $x_{1}=2$, we observe the
establishment of a robust confined pattern with a regular breathing shape
and a gradually growing amplitude (although its spatial extent appears to be
slightly decreasing). For the same management format (\ref{eq:gammaKM1}),
but with $x_{1}=15$, the simulations produce a persistent confined pattern
(within the region of action of the management)
with a more complex structure. Its distinct feature is the presence of
individual large amplitude events (within the
domain of focusing nonlinearity) emerging and disappearing in
quick succession in a way reminiscent of PSs.

\begin{figure}[tbp]
\centering
\begin{tabular}{ccc}
\includegraphics[width=4.5cm]{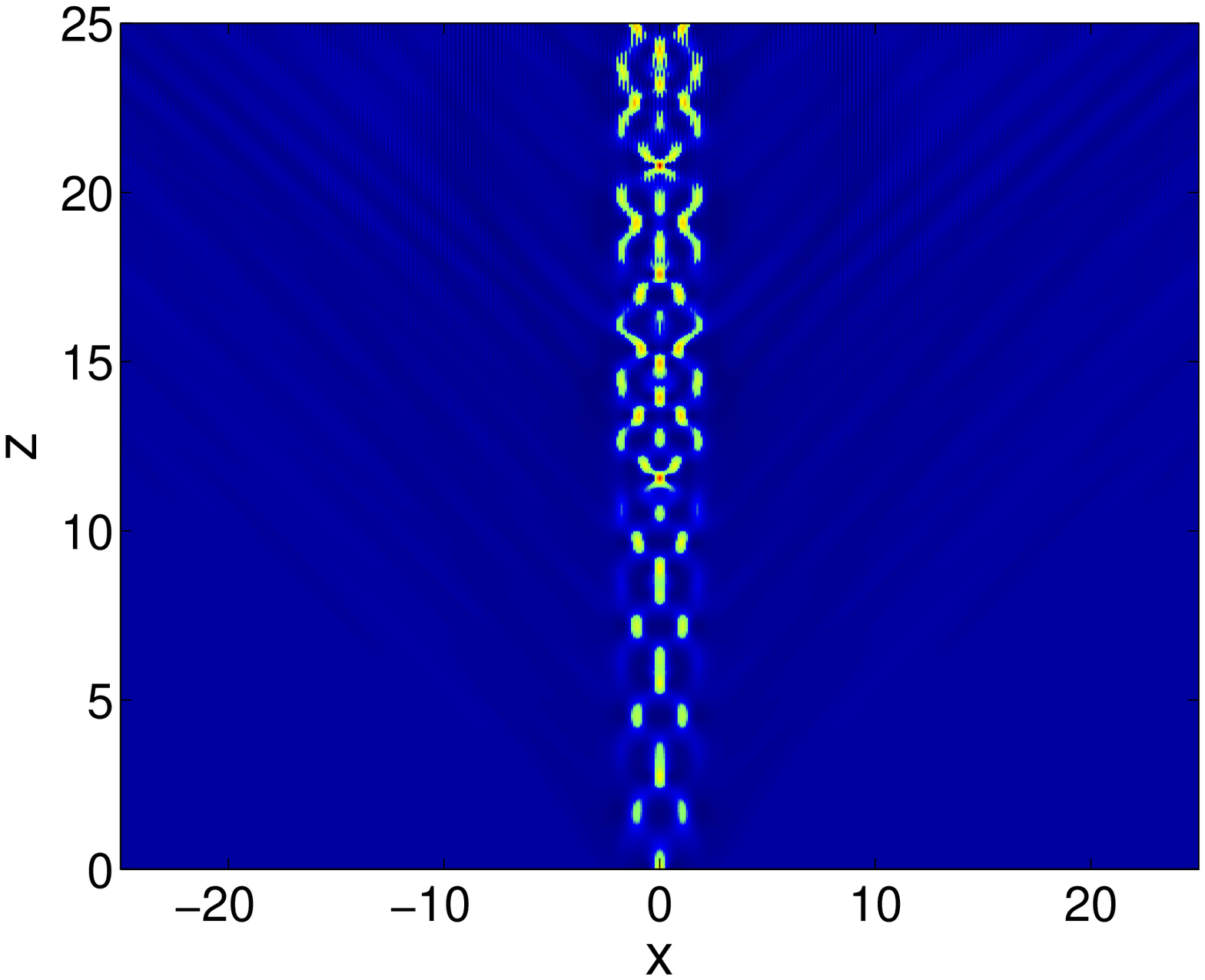} & %
\includegraphics[width=4.5cm]{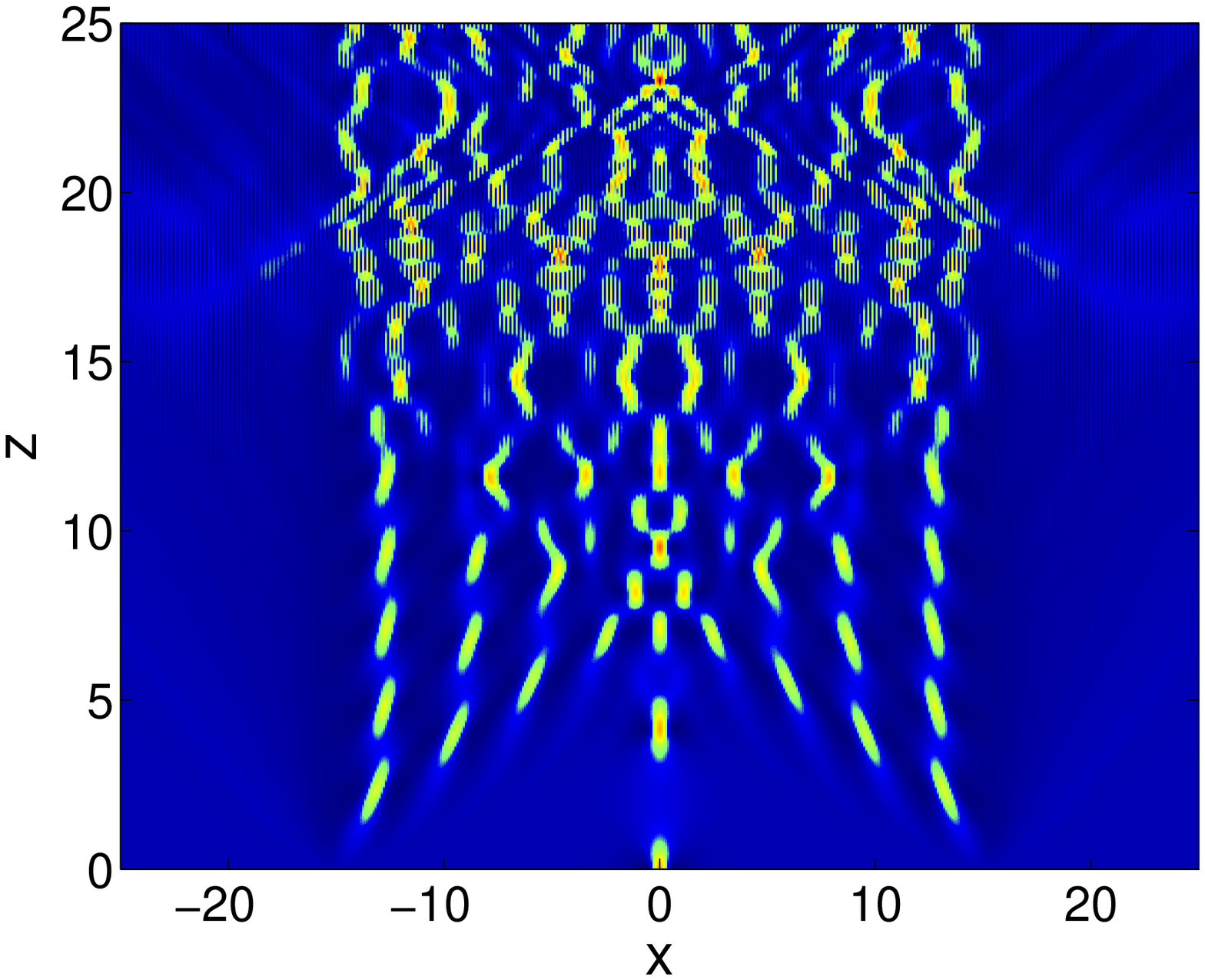} & %
\includegraphics[width=4.5cm]{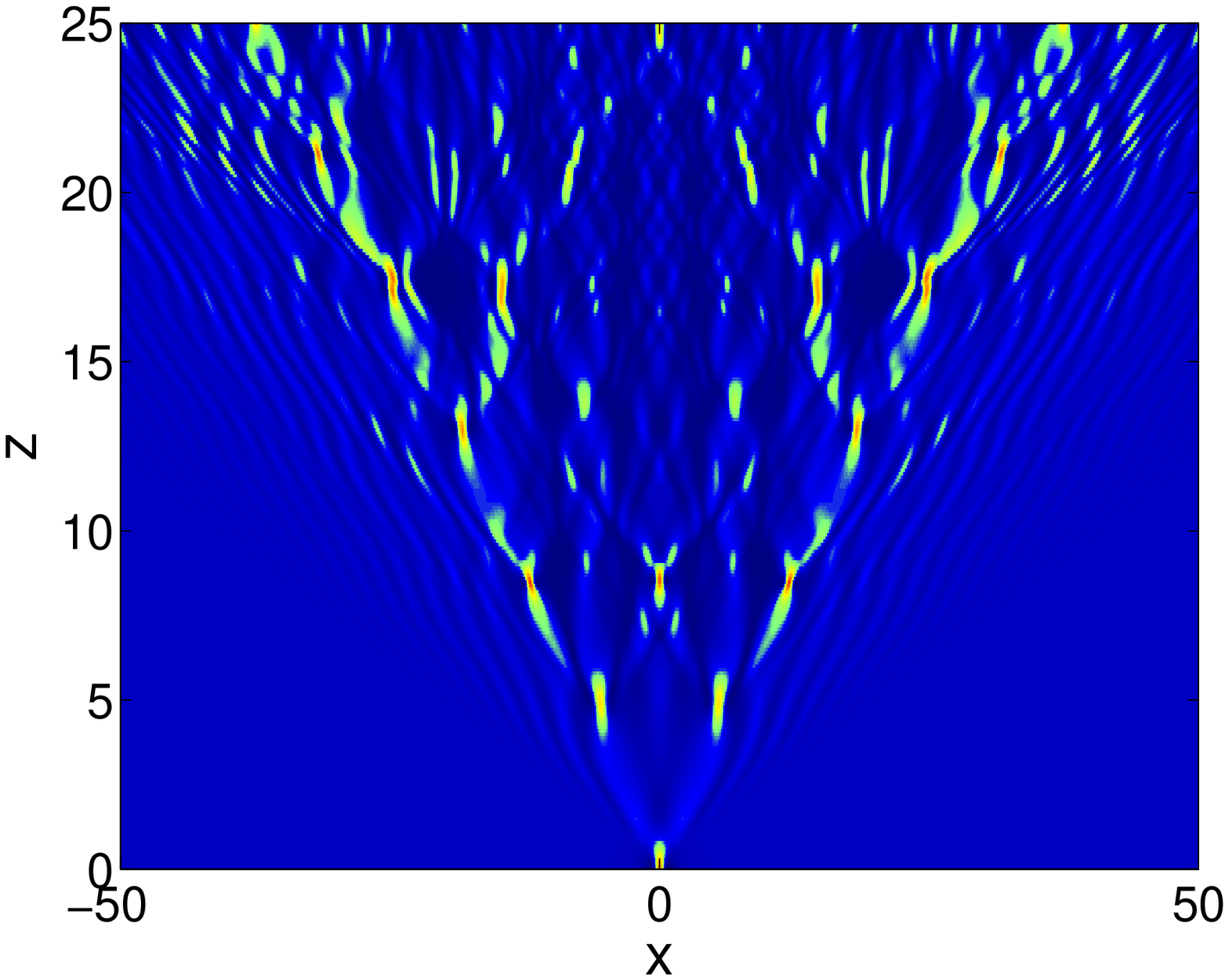} \\
\includegraphics[width=4.5cm]{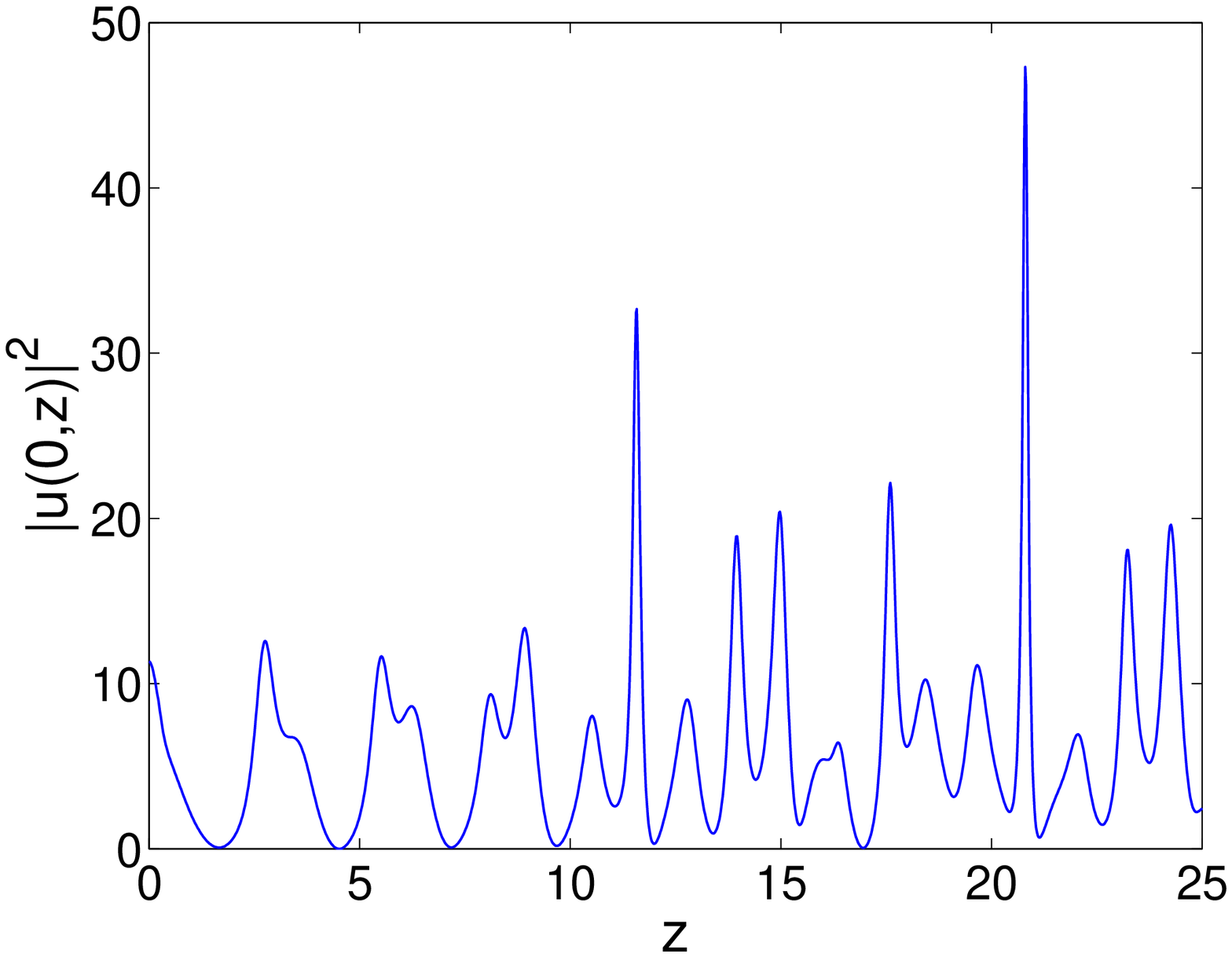} & %
\includegraphics[width=4.5cm]{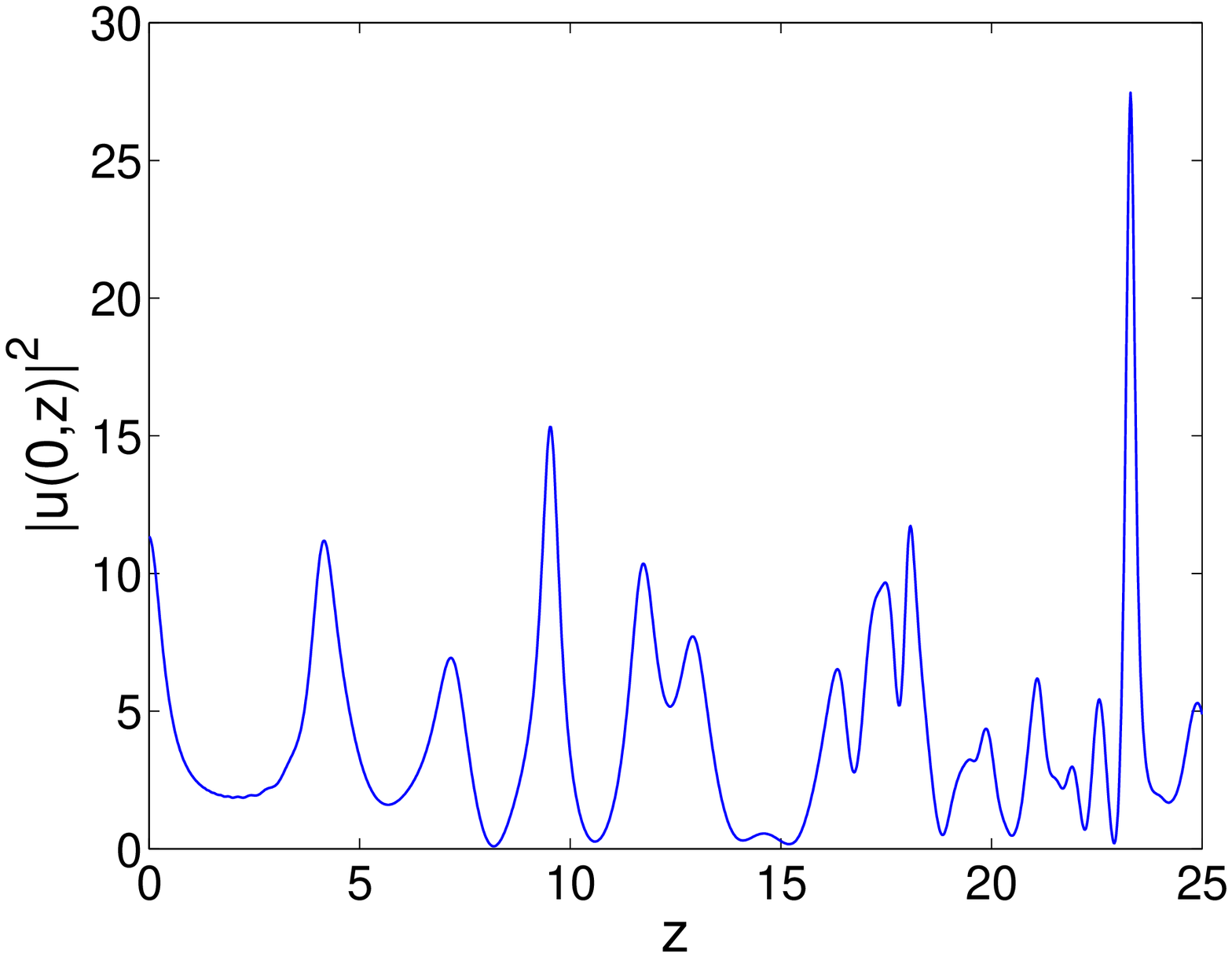} & %
\includegraphics[width=4.5cm]{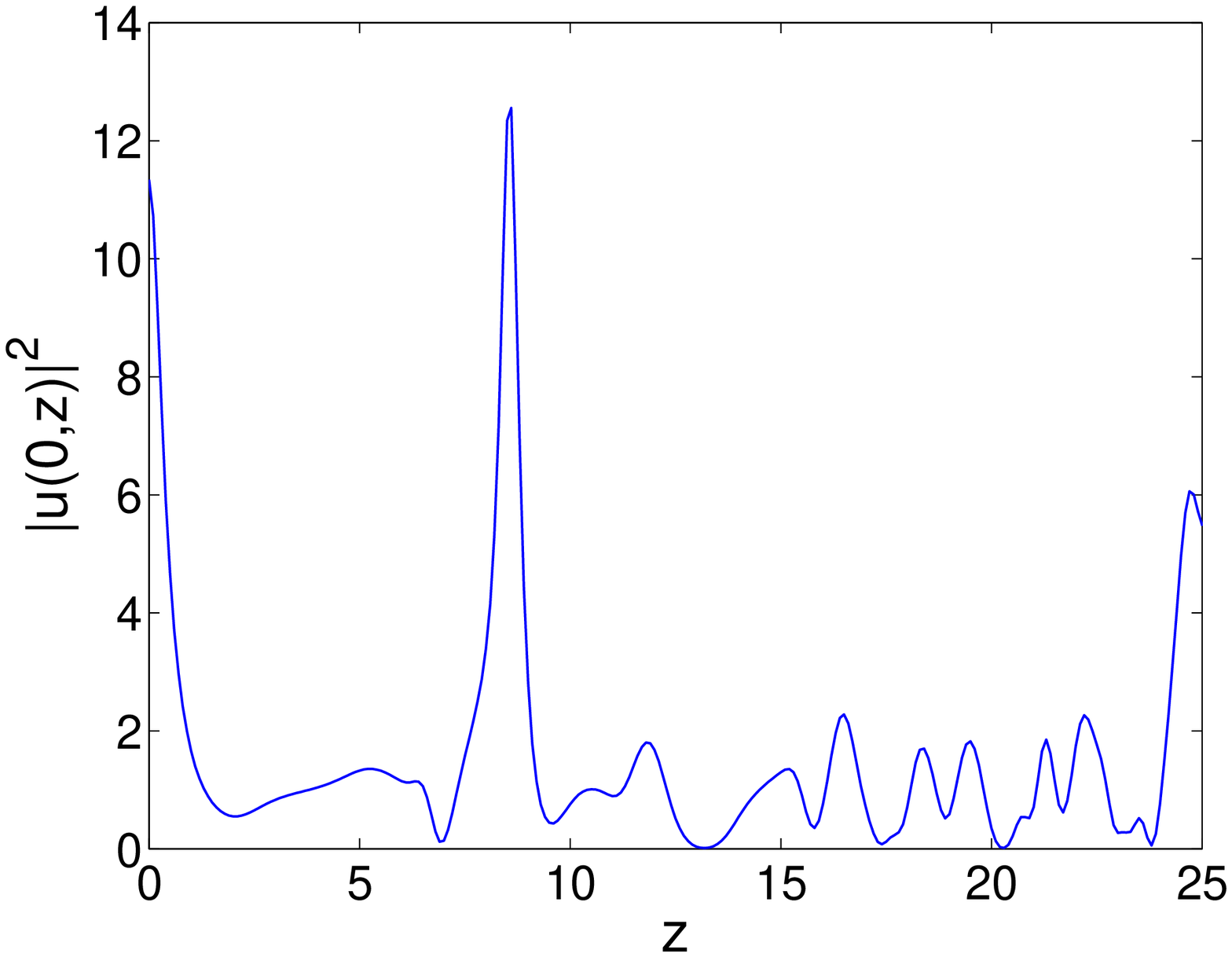} \\
\includegraphics[width=4.5cm]{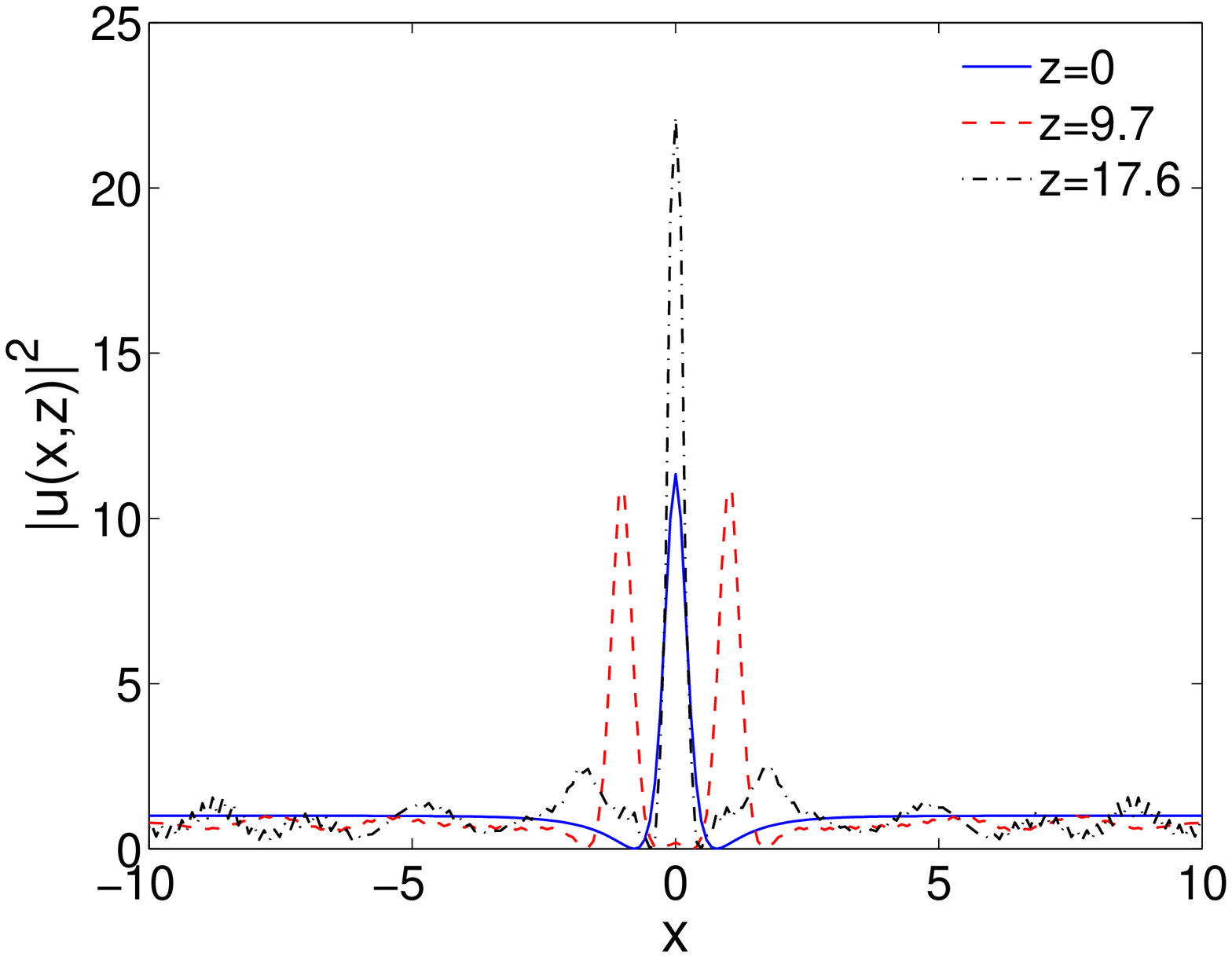} & %
\includegraphics[width=4.5cm]{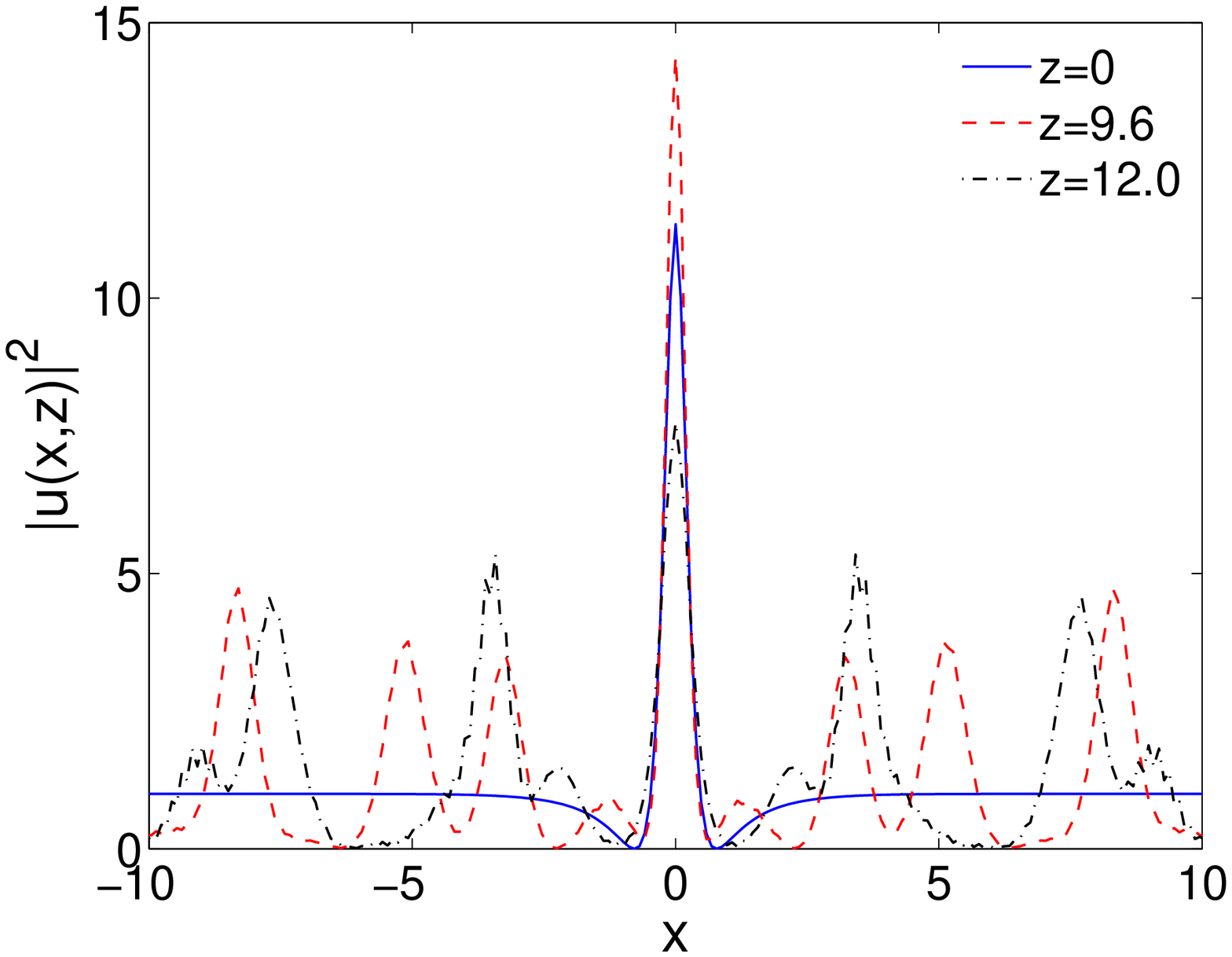} & %
\includegraphics[width=4.5cm]{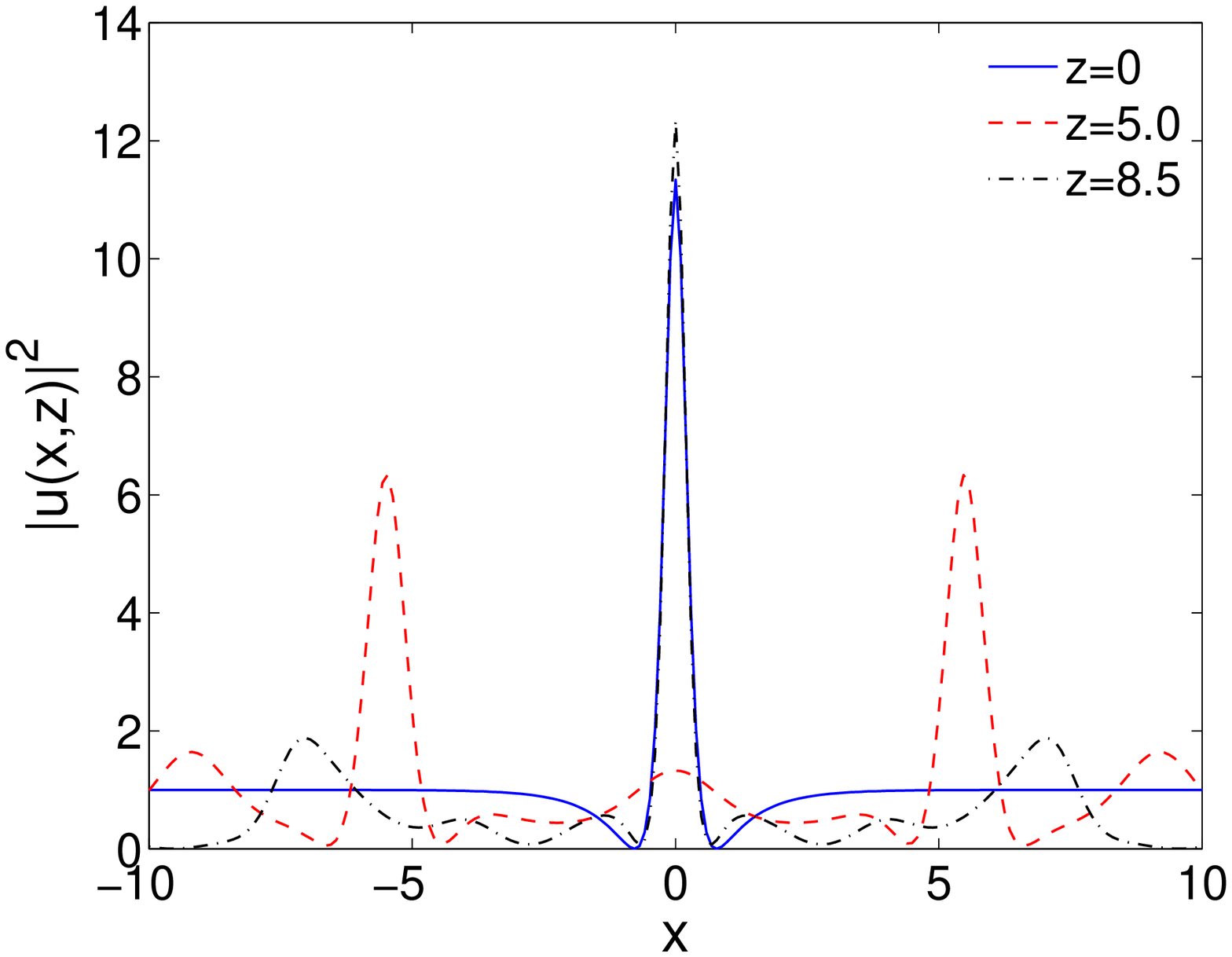} \\
&  &
\end{tabular}%
\caption{Evolution of Kuznetsov-Ma breathers under the action of the
nonlinearity management in the format (\protect\ref{eq:gammaKM1}), for $x_{1}=2
$ and $x_{1}=15$ (left and central columns, respectively), and in the
format (\protect\ref{eq:gammaKM2}) (right column). The top panels display
density plots of the breathers. The middle panels represent the evolution of the
density at $x=0$. The bottom panels show profiles of the breathers at different
values of $z$.}
\label{fig:KMmanaged}
\end{figure}

The $z$-periodic management format (\ref{eq:gammaKM2}) produces a completely
different picture, as seen in the right column of Fig. \ref{fig:KMmanaged}:
the seed breather undergoes initial splitting, which is followed by a
cascade of splittings, and systematic expansion of the area occupied by the
multi-breather pattern. Here, the large amplitude events are less
``ordered'' in their emergence (and are less transparently
persistent at $x=0$), yet they still appear to be present in the
short-intermediate scale dynamics monitored herein.

\section{Conclusion}

In conclusion, we demonstrated the possibility to make
rogue waves (RWs) stable objects in NLSE models, avoiding the modulational
instability of the backgrounds on top of which they arise.
This was achieved by applying the appropriately designed schemes
of the dispersion and nonlinearity management to the CW background
supporting the RWs in the form of the Peregrine soliton (PS) and
Kuznetsov-Ma breathers (KMBs). In particular, it was found that both types of
management, applied along the propagation distance, indeed stabilize the PS,
generating, after its disappearance, additional dynamically persistent
features, in the form of dispersive shock waves and, sometimes, an
additional pair of slowly separating dark solitons. On the other hand, the
nonlinearity management, which makes the NLSE defocusing outside of a finite
domain in the transverse direction, stabilizes the KMBs in the form of
robustly propagating confined breather-like states, while the nonlinearity
management applied periodically along the propagation direction gives rise
to expanding patterns driven by cascading fissions of the breathers.

As further development of the analysis, it may be interesting to consider
interactions of two or several PSs in the framework of the present models,
based on the dispersion and nonlinearity management. 
Other extensions of this work include the investigation of
interactions of PSs with defects or their consideration
in higher dimensions (under stabilized backgrounds).
These directions are presently
under consideration and will be reported accordingly in future studies.

\vspace{+10pt}

\textbf{Acknowledgements.}

J. C.-M. thanks financial support from MAT2016-79866-R project
(AEI/FEDER,UE). P.G.K. gratefully acknowledges the support of NSF-PHY-1602994, the
Alexander von Humboldt Foundation, the Stavros Niarchos Foundation via the
Greek Diaspora Fellowship Program, and the ERC under FP7, Marie Curie
Actions, People, International Research Staff Exchange Scheme
(IRSES-605096). The work of B.A.M. is supported, in part, by the joint
program in physics between NSF and Binational (US-Israel) Science Foundation
through project No. 2015616, and by the Israel Science Foundation through
Grant No. 1286/17. This author also appreciates a grant provided by the
European Erasmus Plus program for visiting the National and Kapodistrian
University of Athens (Greece). P.G.K and D.J.F. acknowledge that this work made possible by NPRP grant \#[8-764-160] from Qatar
National Research Fund (a member of Qatar Foundation). The findings achieved herein are solely the
responsibility of the authors.

\vspace{+10pt}

\end{document}